\documentclass[a4paper,fleqn,usenatbib]{mnras}

\usepackage{newtxtext,newtxmath}

\usepackage[T1]{fontenc}
\usepackage{ae,aecompl}

\usepackage{graphicx}	
\usepackage{amsmath}	
\usepackage{amssymb}	
\usepackage{color}
\usepackage[english]{babel}
\usepackage{graphicx}
\usepackage{hyperref}
\usepackage{listings}
\usepackage{lscape}
\usepackage{natbib}
\usepackage{url}
\usepackage{breakurl}
\usepackage{xspace}
\usepackage{placeins}
\bibpunct{(}{)}{;}{a}{}{,}

\newcounter{Rco}
\newcommand{\object}[1]{#1}

\newcommand{\Ionst}[1]{\setcounter{Rco}{#1}\Roman{Rco}}
\newcommand{\Ion}[2]{\mbox{#1\,{\scriptsize\Ionst{#2}}}}
\newcommand{\Ionw}[3]{\mbox{#1\,{\scriptsize\Ionst{#2}}~$\lambda\,#3$\,\AA}\xspace}

\newcommand{\Ionww}[3]{\mbox{#1\,{\scriptsize\Ionst{#2}}~$\lambda\lambda\,#3$\,\AA}\xspace}
\newcommand{\Jonw}[3]{\mbox{\ion{#1}{#2}~$\lambda\,#3$\,\AA}\xspace}
\newcommand{\Jonww}[3]{\mbox{\ion{#1}{#2}~$\lambda\lambda\,#3$\,\AA}\xspace}

\newcommand{\logg}{\mbox{$\log g$}\xspace}
\newcommand{\loggw}[1]{\mbox{$\log g\hspace{-0.5mm} =\hspace{-0.5mm}  #1$}}

\newcommand{\ab}[1]{\mbox{Fig.\,\ref{#1}}}
\newcommand{\sA}[1]{\mbox{(Fig.\,\ref{#1})}}

\newcommand{\sK}[1]{\mbox{(Sect.\,\ref{#1})}}

\newcommand{\ta}[1]{\mbox{Table\,\ref{#1}}}
\newcommand{\sT}[1]{\mbox{(Table\,\ref{#1})}}
\newcommand{\Teff}{\mbox{$T_\mathrm{eff}$}\xspace}

\newcommand{\Teffw}[1]{\mbox{$\Teff\hspace{-0.5mm} =\hspace{-0.5mm} #1 \,\mathrm{K}$}}
\newcommand{\ebv}{\mbox{$E_\mathrm{B-V}$}}
\newcommand{\ebvw}[1]{\mbox{$\ebv\hspace{-0.5mm} =\hspace{-0.5mm} #1$}}

\newcommand{\pna}{Abell\,43\xspace}
\newcommand{\wda}{WD\,1751+106\xspace}
\newcommand{\pnn}{NGC\,7094\xspace}
\newcommand{\wdn}{WD\,2134+125\xspace}

\title[Analysis of the CSPNe \pna and \pnn]{Spectral analysis of the hybrid PG\,1159-type central stars \\
       of the planetary nebulae \pna and \pnn
\thanks{Based on observations with the NASA/ESA Hubble Space Telescope, obtained at the Space Telescope Science 
           Institute, which is operated by the Association of Universities for Research in Astronomy, Inc., under 
           NASA contract NAS5-26666.}
          \thanks{Based on observations made with the NASA-CNES-CSA Far Ultraviolet Spectroscopic Explorer.}
          \thanks{Based on data products from observations made with ESO Telescopes at the La Silla Paranal Observatory under programme ID 167.D-0407.}}

\author[L\@. L\"obling et al.]{L\@. L\"obling$^{1}$\thanks{E-mail: loebling@astro.uni-tuebingen.de},
T\@. Rauch$^{1}$, 
M\@. M\@. Miller Bertolami$^{2,3}$, 
H\@. Todt$^{4}$, 
F\@. Friederich$^{1}$, \newauthor
M\@. Ziegler$^{1}$,
K\@. Werner$^{1}$
and J\@. W\@. Kruk$^{5}$ \\
$^{1}$Institute for Astronomy and Astrophysics, Kepler Center for Astro and Particle Physics,
Eberhard Karls University, \\Sand 1, 72076 T\"ubingen, Germany \\
$^{2}$Instituto de Astrof\'{i}sica La Plata, CONICET-UNLP, Paseo del Bosque s/n, (B1900FWA) La Plata, Argentina\\
$^{3}$Facultad de Ciencias Astron\'{o}micas y Geof\'{i}sicas, UNLP,  Paseo del Bosque s/n, (B1900FWA) La Plata, Argentina\\
$^{4}$Institute of Physics and Astronomy, University of Potsdam, Karl-Liebknecht-Str\@. 24/25, 14476 Potsdam, Germany\\
$^{5}$NASA Goddard Space Flight Center, Greenbelt, MD\,20771, USA           
}

\date{Accepted 2019 July 16. Received 2019 June 18; in original form 2019 April 30}

\pubyear{2019}

\begin{document}
\label{firstpage}
\pagerange{\pageref{firstpage}--\pageref{lastpage}}
\maketitle

\begin{abstract}
Stellar post asymptotic giant branch (post-AGB) evolution 
can be completely altered by a
final thermal pulse (FTP) which may occur 
when the star is still leaving the AGB (AFTP), at the departure from the AGB at
still constant luminosity (late TP, LTP) or 
after the entry to the white-dwarf cooling sequence (very late TP, VLTP). 
Then convection mixes the He-rich material with the H-rich envelope.
According to stellar evolution models the
result is a star with a surface composition of $\mathrm{H}\approx\,20$\,\% by mass (AFTP), 
$\approx 1$\,\%  (LTP), or (almost) no H (VLTP).
Since FTP stars exhibit intershell material at their surface, spectral analyses
establish constraints for AGB nucleosynthesis and stellar evolution.
We performed a spectral analysis of the so-called hybrid PG\,1159-type central stars (CS)
of the planetary nebulae \pna and \pnn
by means of non-local thermodynamical equilibrium models.
We confirm the previously determined
effective temperatures of \Teffw{115\,000\pm 5\,000} and determine
surface gravities of $\log (g\,/\,\mathrm{cm/s^2}) = 5.6\pm 0.1$
for both.
From a comparison with AFTP evolutionary tracks, we derive stellar masses of
$0.57^{+0.07}_{-0.04}$\,$M_\odot$ and determine the abundances of H, He, and metals up to Xe.
\textcolor{black}{Both CS are likely AFTP stars with a surface H mass fraction of $0.25 \pm 0.03$ and $0.15 \pm 0.03$, respectively, and
a Fe deficiency indicating subsolar initial metallicities. The light metals show typical PG\,1159-type abundances and the elemental composition is in good agreement with predictions from AFTP evolutionary models. However, the expansion ages do not agree with evolution timescales expected from the AFTP scenario and alternatives should be explored.}
\end{abstract}




\section{Introduction}
\label{sect:intro}

\begin{figure*} 
  \resizebox{\hsize}{!}{\includegraphics{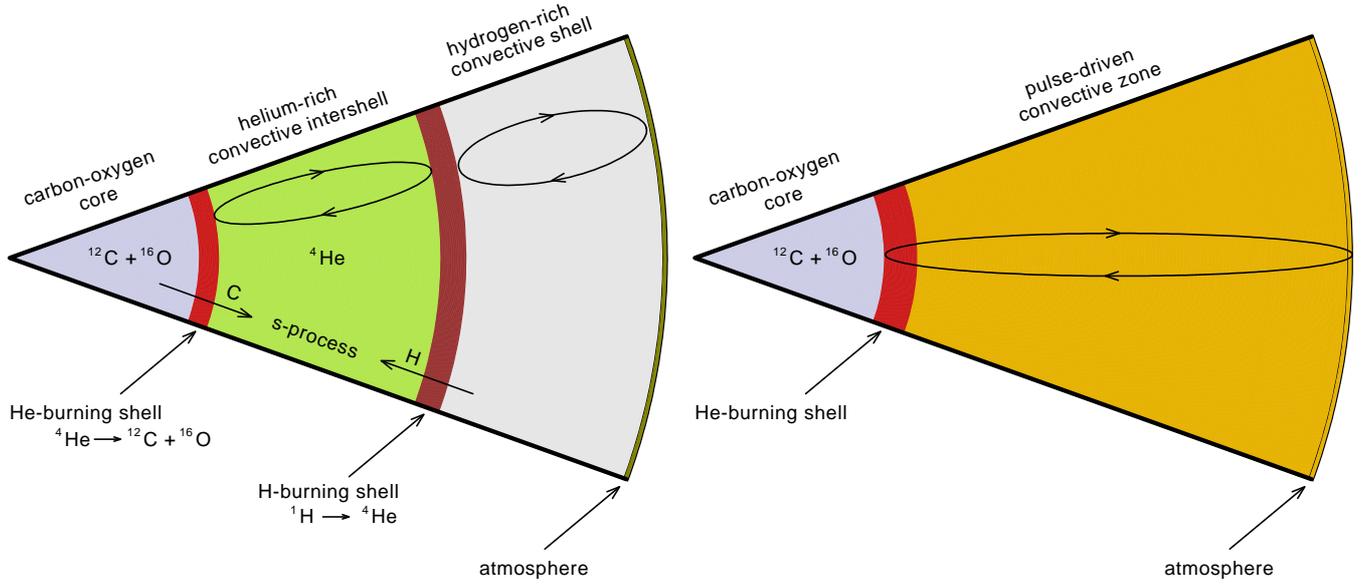}}
\caption{Left: Internal structure (not to scale) of an AGB star during a thermal pulse. Right: Internal structure of a post-AGB star after having experienced a final thermal pulse that caused flash induced mixing of the envelope and the intershell region. }
   \label{fig:internalstructure}
\end{figure*}
Asymptotic giant branch (AGB) stars are  important contributors to the formation of elements heavier than iron (trans-iron elements, TIEs). 
Schematically, the internal structure of an AGB star is illustrated in \ab{fig:internalstructure}. It is composed of an inner C/O core, the two burning shells with a He, C, and O rich intershell region in between and an H-rich convective envelope on top. These stars experience several thermal pulses (TPs) during which the intershell region becomes convectively unstable and C-rich both due to He burning and to dredge up from the core. Additionally, small amounts of H can be partially mixed into the intershell region during the expansion and cooling of the envelope that follows a TP. The presence of large amounts of $^{12}$C mixed with traces of H at high temperatures leads to the formation of $^{13}$C that acts as a neutron source for the  slow neutron capture process (s-process). 
The intershell region of AGB stars is the main astrophysical site for the s-process.
The stellar post-AGB evolution divides into two major
channels of H-rich and H-deficient stars. The latter comprise about
a quarter of all post-AGB stars and include He- and C-dominated stars.
While the He-dominated, H-deficient stars may be the result of stellar mergers \citep{reindletal2014},
it is commonly accepted that the C-rich are the outcome of a (very) late He-shell flash 
\citep[late thermal pulse, LTP, cf\@.][]{wernerherwig2006}.
The occurrence of a thermal pulse in a post-AGB star or white dwarf was predicted by, e.g., \citet{paczynski1970}, \citet{schoenberner1979}, and \citet{ibenetal1983}.
The particular timing of the final thermal pulse (FTP), determines the amount of remaining photospheric H 
\citep[cf.,][]{herwig2001}. Still on the AGB (AGB final thermal pulse, AFTP), flash-induced mixing of the H-rich
envelope ($\approx 10^{-2}\,M_\odot$) with the He-rich intershell layer ($\approx 10^{-2}\,M_\odot$)
reduces the H abundance to about $10 - 20$\,\% but \ion{H}{i} lines remain detectable. After the departure from the
AGB, the H-rich envelope is less massive ($10^{-4}\,M_\odot$). If the nuclear
burning is still ``on'', i.e\@., the star evolves at constant luminosity, the mixing due to a late thermal pulse (LTP)
reduces H below the detection limit (about 10\,\% by mass at the relatively high surface gravity). 
After the star has entered the white-dwarf cooling sequence and the nuclear burning is ``off'', 
a very late thermal pulse (VLTP) will produce convective mixing of the entire H-rich envelope 
(no entropy barrier due to the H-burning shell) down to the bottom of the He-burning shell where
H is burned. In that case, the star will become H free at that time.
The internal structure of such a post-AGB star that underwent a FTP scenario is illustrated in \ab{fig:internalstructure}.\\
The spectroscopic class of PG\,1159 stars 
(effective temperatures of $75\,000\,\mathrm{K} \la \Teff \la 250\,000\,\mathrm{K}$ and
surface gravities of $5.5 \la \log (g\,/\,\mathrm{cm/s^2}) \la 8.0$)
belongs to the H-deficient, C-rich evolutionary channel \citep[e.g.,][]{wernerherwig2006}, with the sequence
AGB $\rightarrow$ [WC]-type Wolf-Rayet stars  $\rightarrow$ PG\,1159 stars  $\rightarrow$ DO-type white dwarfs (WDs). 
In the AFTP and LTP scenarios with any remaining H, the stars will turn into DA-type WDs.
In PG\,1159 star photospheres, He, C, and O are dominant with mass fractions of 
He = [0.30,0.92], C = [0.08,0.60], and O = [0.02,0.20]
\citep{werneretal2016}.\\
\citet{napiwotzkischoenberner1991} discovered the spectroscopic sub-class of so-called hybrid PG\,1159 stars.
They found that 
\object{WD\,1822+008} \citep{mccooksion1999,mccooksion1999cat}, the central star (CS) of the planetary nebula (PN)  
\object{Sh\,2-68} exhibits
strong Balmer lines in its spectrum. 
The hybrid PG\,1159 stars are thought to be AFTP stars.
Presently, only five of them are known, namely the 
CSPNe of 
\pna, 
\pnn,
\object{Sh\,2$-$68}, 
\object{HS\,2324+3944}, and
\object{SDSS\,152116.00+251437.46} \citep{wernerherwig2006,werneretal2014}.\\
\pna (\object{PN\,G036.0+17.6}) 
was discovered by \citet[][object No\@. 31]{abell1955} and 
classified as PN \citep[][No\@. 43]{abell1966}.
\pnn (\object{PN\,G066.7$-$28.2}) was discovered in 1885 by \citet{swift1885}. 
\citet{kohoutek1963} identified it as a PN (\object{K\,1$-$19}). 
Narrow-band imaging of \pna and \pnn \citep{rauch1999} revealed apparent sizes 
(in West-East and North-South direction) of 
1\arcmin 28\arcsec\,$\times$\,1\arcmin 20\arcsec  and
1\arcmin 45\arcsec\,$\times$\,1\arcmin 46\arcsec, respectively.\\
\pna and \pnn belong to the group of so-called ``Galactic Soccerballs'' \citep{rauch1999} because they
exhibit filamentary structures that remind of the seams of a traditional leather soccer ball.
These structures may be explained by instabilities in the dense, moving nebular shell \citep{vishniac1983}. While \pna is almost perfectly round and most likely expanded into a void in the ISM, \pnn shows some deformation that may be a hint for ISM interaction.
Another recently discovered member of this group is the 
\object{PN\,Kn\,61} \citep[\object{SDSS\,J192138.93+381857.2},][]{kronbergeretal2012,deMarcoetal2015}.
\citet{garciadiazetal2014} compared medium-resolution optical spectra of the CSPN\,Kn\,61 with
spectra published by \citet{werneretal2014} and found a particularly close resemblance of the CSPN\,Kn\,61
to \object{SDSS\,075415.12+085232.18}, an H-deficient PG\,1159-type star with
\Teffw{120\,000 \pm 10\,000},
\loggw{7.0 \pm 0.3},
and a mass ratio C/He = 1.
Other members and candidates to become a Galactic Soccerball nebula are known, e.g.,
the PN \object{NGC\,1501} (\object{PN\,G144.5+06.5}).
An investigation on the 3-dimensional ionization structure by \citet{ragazzonietal2001} had shown that 
it might resemble a Soccerball nebula in a couple of thousands of years. The CSPN, however, is of spectral type 
[WC4]  \citep{koesterkehamann1997} and cannot resemble a progenitor star of the CSs of \pna and \pnn.
In this paper, we analyze the hybrid PG\,1159-type CSs of \pna and \pnn, that we introduce briefly in the following paragraphs.\\
A first spectral analysis of the CSs of \pna and \pnn,
namely \wda and \wdn \citep{mccooksion1999,mccooksion1999cat}, respectively, 
with non-local thermodynamical equilibrium (NLTE) model atmospheres that considered opacities of H, He, and C was presented by \citet{dreizleretal1995}.
They analyzed medium-resolution optical spectra and found that their synthetic spectra,
calculated with
\Teffw{110\,000}, 
\loggw{5.7}, and 
a surface-abundance pattern of H/He/C = 42/51/5 (by mass, H is uncertain),
reproduced equally good the observations of both stars making them a pair
of spectroscopic twins.\\
\citet{napiwotzki1999} used medium-resolution optical spectra and an extended H+He-composed NLTE model-atmosphere
grid. With a statistical ($\chi^2$) approach, he found
\Teffw{116\,900 \pm 5500} and \loggw{5.51 \pm 0.22} for \wda and
\Teffw{125\,900 \pm 7700} and \loggw{5.45 \pm 0.23} for \wdn. An attempt to measure the Fe abundance of \wdn from far ultraviolet (FUV)
observations performed with Far Ultraviolet Spectroscopic Explorer (FUSE) revealed a strong Fe underabundance 
of 1-2\,dex \citep{miksaetal2002}. This was not in line with expectations from stellar evolution theory \citep[e.g.,][]{bussoetal1999}. \citet{ziegleretal2009a} found also an underabundance of Ni of about 1\,dex for both stars. 
The transformation of Fe to Ni seems therefore unlikely to be the reason for the Fe deficiency. 
They reanalyzed \Teff and \logg of \wdn and found \Teffw{100\,000 \pm 15\,000} and \loggw{5.5 \pm 0.2} with an improved 
abundance ratio of H/He = 17/69 (by mass). Furthermore, the element abundances of the C\,--\,Ne, Si, P, and S were determined. 
\citet{ringatetal2011} reanalyzed \wda and found \Teffw{105\,000 \pm 10\,000} and \loggw{5.6 \pm 0.3}. 
Also the element abundances of C\,--\,Ne, Si, P, and S were determined and agree with the values of \citet{friederich2010}.
\citet{loebling2018} found \Teffw{115\,000 \pm 5000} for both stars and \loggw{5.4 \pm 0.1} and \loggw{5.5 \pm 0.1} for \wdn and \wda, respectively. She considered 31 elements in her analysis and determined abundances in individual line-formation calculations. The present work is a continuative analysis giving a more extensive description. \\
For NLTE model-atmosphere calculations, reliable atomic data is mandatory to construct detailed model
atoms to represent individual elements. In the last decade, the availability of such atomic data
improved, e.g., Kurucz's line lists for iron-group elements (IGEs), namely Ca - Ni, were strongly extended in 2009 
\citep{kurucz2009,kurucz2011} by about a factor of ten. 
In addition, transition probabilities and oscillator strengths for many TIEs were calculated
recently (Table\,\ref{tab:tee}). 
Therefore, we decided to perform
a detailed spectral analysis of the hybrid PG\,1159-type CSPNe \pna and \pnn,
by means of state-of-the-art NLTE model-atmosphere techniques. 
We describe the available observations and our model atmospheres in Sects.\,\ref{sect:obs} and \ref{sect:model}, respectively.
The spectral analyses follow in Sects\@.\,\ref{sect:tefflogg} and \ref{sect:metal}.
We investigate on the stellar wind of both stars in Sect\@.\,\ref{sect:powr} and determine
stellar masses, distances, and luminosities in Sect\@.\,\ref{sect:mld}.
We summarize the results and conclude in Sect\@.\,\ref{sect:results}.

\section{Observations}
\label{sect:obs}
Our spectral analysis is based on high signal-to-noise ratio ($S/N$) and high-resolution observations from the far ultraviolet (FUV) to the optical wavelength range. UV spectra were retrieved from 
the Barbara A\@. Mikulski Archive for Space Telescopes (MAST).
To improve the $S/N$, multiple observations in the same setup were co-added.
The spectra were partly processed with a low-pass filter \citep{savitzkygolay1964}.
To simulate the instruments' resolutions, all synthetic spectra shown in this paper are convolved 
with respective Gaussians. \textcolor{black}{The observation log for all space and ground based observations of \wda and \wdn used for this work is given in \ta{tab:obslog}.}

\paragraph*{Radial and rotational velocity.}
\textcolor{black}{We measured radial velocity shifts for all observations using prominent lines of \Ion{He}{2}, \Ion{C}{4}, \Ion{O}{5} and \textsc{vi}, \Ion{Si}{5}, and \Ion{Fe}{7} and shifted \textcolor{black}{the spectra} to rest wavelength.\\
The observed line profiles are broadened but the quality of the spectra does not unambiguously allow to decide whether it is due to stellar rotation or caused by some wind related macro turbulence. For \wda, we selected \Ionww{O}{6}{1124.7, 1124.9} and \Ionw{S}{6}{1117.8}
 \sA{fig:rotation} to determine
a rotational velocity of $v_\mathrm{rot} \sin i = 18 \pm 5\,\mathrm{km}/\mathrm{s}$. This new determination revises the previous higher value of $v_\mathrm{rot} \sin i = 42 \pm 13\,\mathrm{km}/\mathrm{s}$
\citep{rauchetal2004}. The profiles of these lines also agree with broadening with radial-tangential macro turbulence profiles \citep{gray1975} with the same velocity \sA{fig:rotation}. 
For \wdn, we used \Ion{O}{6} $\lambda\lambda\,1122.4, 1122.6, 1124.7, 1124.9$ and \Ionw{N}{5}{1242.6} \sA{fig:rotation} to  determine $v_\mathrm{rot} \sin i = 28 \pm 5\,\mathrm{km}/\mathrm{s}$. This value agrees within the error limits with the value of $46 \pm 16\,\mathrm{km}/\mathrm{s}$
from \citet{rauchetal2004}. Again, we cannot claim this broadening to be rotation alone because the profiles can also be reporduced with radial-tangential macro turbulence profiles with $v_\mathrm{macro} = 35 \pm 5\,\mathrm{km}/\mathrm{s}$.}\\
\begin{figure*} 
  \resizebox{\hsize}{!}{\includegraphics{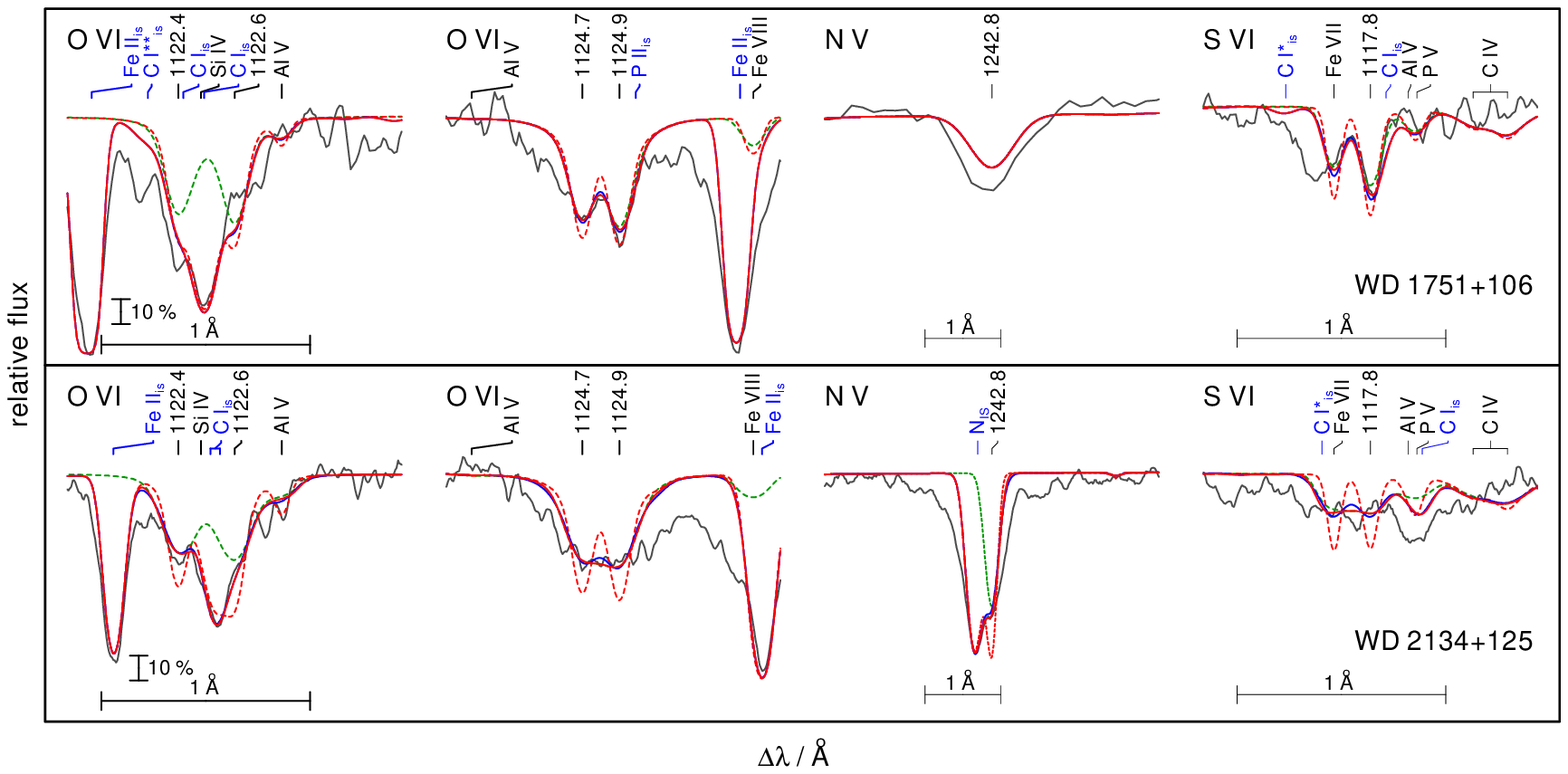}}
   \caption{Synthetic spectra (red) convolved with rotational profiles ($v_\mathrm{rot} = 18\,\mathrm{km}/\mathrm{s}$ for \wda, upper panels and $v_\mathrm{rot} = 28\,\mathrm{km}/\mathrm{s}$ for \wdn, lower panels) and convolved with radial-tangential macro turbulence profiles (blue, $v_\mathrm{macro}=18\,\mathrm{km}/\mathrm{s}$ for \wda, upper panels and $v_\mathrm{macro} = 35\,\mathrm{km}/\mathrm{s}$ for \wdn, lower panels) around 
            \ion{O}{vi}, \ion{N}{v}, and \ion{S}{vi} lines (marked with their wavelengths at
            the top of the panels)
            calculated from our final models compared with observations (gray).
            A model without extra broadening (red, dashed) 
            and one without interstellar absorption (green, dashed) are shown.
            Interstellar absorption lines are indicated by blue marks.}
   \label{fig:rotation}
\end{figure*}
\paragraph*{Interstellar reddening} was measured by a comparison of observed UV fluxes and
optical and infrared brightnesses with our synthetic spectra. The latter were normalized 
to the Two Micron All Sky Survey \citep[2MASS,][]{skrutskieetal2006,cutrietal2003} brightnesses
and then, Fitzpatrick's law \citep{fitzpatrick1999} was applied to match the observed UV 
continuum flux level (Fig. \ref{fig:ebv}). 
We determined
$\ebv = 0.265 \pm 0.01$ and
$\ebv = 0.135 \pm 0.01$ for \wda and \wdn, respectively.\\
We determined the interstellar neutral H column density from the comparison of theoretical
line profiles of Ly\,$\alpha$ with the observations (Fig. \ref{fig:nh}). These are best 
reproduced at 
$n_\ion{H}{i}= 1.0 \pm 0.1 \times 10^{21}\,\mathrm{cm^{-2}}$ and
$n_\ion{H}{i}= 6.5 \pm 0.1 \times 10^{20}\,\mathrm{cm^{-2}}$ for \wda and \wdn, respectively.
Our values of 
$\log(n_\mathrm{\ion{H}{i}}/\ebv) = 21.58 \pm 0.02$ and 
$\log(n_\mathrm{\ion{H}{i}}/\ebv) = 21.68 \pm 0.03$, respectively,
agree well with the prediction from the Galactic reddening law of 
\citet[][$\log(n_\mathrm{\ion{H}{i}}/\ebv) = 21.58\pm 0.1$]{groenewegenlamers1989}.

\paragraph*{Interstellar line absorption} in the FUSE observations was modeled with the line-profile fitting procedure OWENS 
\citep{lemoineetal2002,hebrardetal2002,hebrardetal2003}.
It allows to consider several individual clouds in the interstellar medium (ISM) with individual chemical compositions, 
column densities for each of the included molecules and ions, radial and turbulent velocities, and temperatures.  
The FUV observations are strongly contaminated by ISM line absorption and, thus, it is necessary to reproduce 
these lines well to unambiguously identify photospheric lines \citep[cf\@.,][]{ziegleretal2007,ziegleretal2012}.
In the FUSE spectra of \wda and \wdn, 
ISM absorption lines from 
$\mathrm{H}_2$ ($\mathrm{J}=0-5$), 
HD,
\ion{C}{i-iii}, 
\ion{N}{i-ii}, 
\ion{O}{i}, 
\ion{Si}{ii}, 
\ion{P}{ii}, 
\ion{S}{iii}, 
\ion{Ar}{i}, and
\ion{Fe}{ii} 
were identified and simulated.  

\section{Model atmospheres and atomic data}
\label{sect:model}

To calculate synthetic spectra, we used the T\"ubingen NLTE
Model Atmosphere Package 
\citep[TMAP\footnote{\url{http://astro.uni-tuebingen.de/~TMAP}},][]{werneretal2003a,tmap2012}.
The models assume plane-parallel geometry, are chemically homogeneous, and in hydrostatic and radiative 
equilibrium. TMAP considers level dissolution (pressure ionization) following
\citet{hummermihalas1988} and \citet{hubenyetal1994}. 
Stark-broadening tables of 
\citet[][extended tables of 2015, priv\@. comm\@.]{tremblaybergeron2009} and 
\citet{schoeningbutler1989}
are used to calculate the theoretical profiles of \ion{H}{i} and \ion{He}{ii} lines, respectively.
To re\-present the elements considered by TMAP, model atoms
were retrieved from the T\"ubingen Model Atom Database
\citep[TMAD,][]{rauchdeetjen2003} that
has been constructed as part of the T\"ubingen contribution to the German Astrophysical Virtual Observatory 
(GAVO\footnote{\url{http://www.g-vo.org}}).
For IGEs and TIEs (atomic weight $Z \ge 29$), we used
Kurucz's line lists\footnote{\url{http://kurucz.harvard.edu/atoms.html}} \citep{kurucz2009,kurucz2011}
and recently calculated data for
Zn, Ga, Ge, Se, Kr, Sr, Zr, Mo, Te, I, Xe, and Ba (Table\,\ref{tab:tee})
that is available via the
T\"ubingen Oscillator Strengths Service (TOSS).
For the elements with $Z \ge 20$, we created model atoms using a statistical approach that calculates super
levels and super lines \citep{rauchdeetjen2003}. 
The statistics of all elements considered in our model-atmosphere calculations are summarized in Table\,\ref{tab:stat}.\\
To simulate prominent P\,Cygni profiles in the observations, we used the Potsdam Wolf-Rayet (PoWR) code that has been developed 
for expanding atmospheres and considers mass loss due to a stellar wind (Sect\@.\ref{sect:powr}). These models are used to determine  mass loss rates and terminal wind velocities.

\section{Effective temperature and surface gravity}
\label{sect:tefflogg}

Model atmospheres grids ($\Delta \Teffw{5000}$ and $\Delta \loggw{0.1}$) were calculated around the literature 
values of \citet{loebling2018}.
These models consider opacities of 31 elements from H to Ba for which the ionization fractions of the considered ions are shown in \ab{fig:ionfrac}. \textcolor{black}{The abundances used for the calculation of the atmospheric structure are given in \ta{tab:pro2}}.
The best agreement for the line width and depth increment for the observed
\Jonww{He}{ii}{4025.6, 4100.1, 4199.8, 4338.7, 4859.3, 5411.5}, and 
\Jonww{H}{i}{4101.7, 4340.5, 4861.3} was obtained for surface gravity values of \loggw{5.6\pm 0.1} for \wda \sA{fig:logga43} and \wdn \sA{fig:loggngc}. The value for \wdn is also verified by the depth increment of the \Ion{He}{2} Fowler series \sA{fig:fowlerngc}.
The lower surface gravity values of \citet{loebling2018} were based on model atmospheres assuming the abundances found by \citet{friederich2010}. Using new models with a revised He/H ratio and including opacities of more elements, these previous values appear too low. Higher values for $\log g$ have an impact on the final masses which are lower compared to previous values \sK{sect:mld}. 
We confirm the temperature determination of \Teffw{115\,000\pm 5\,000} for both stars by \citet{loebling2018} by the evaluation of the \Ion{O}{5}/\Ion{O}{6} ionization equilibrium using \Jonw{O}{v}{1371.3, 1506.6, 1506.7, 1506.8, 4930.2, 4930.3} 
and \Jonww{O}{vi}{1124.7, 1124.9, 3811.3, 3834.2} \sA{fig:teffall}. 
We adopt \Teffw{115\,000} and \loggw{5.6} for \wda and \wdn for our further analysis.

\section{Metal abundances}
\label{sect:metal}

\begin{figure*}
  \resizebox{\hsize}{!}{\includegraphics{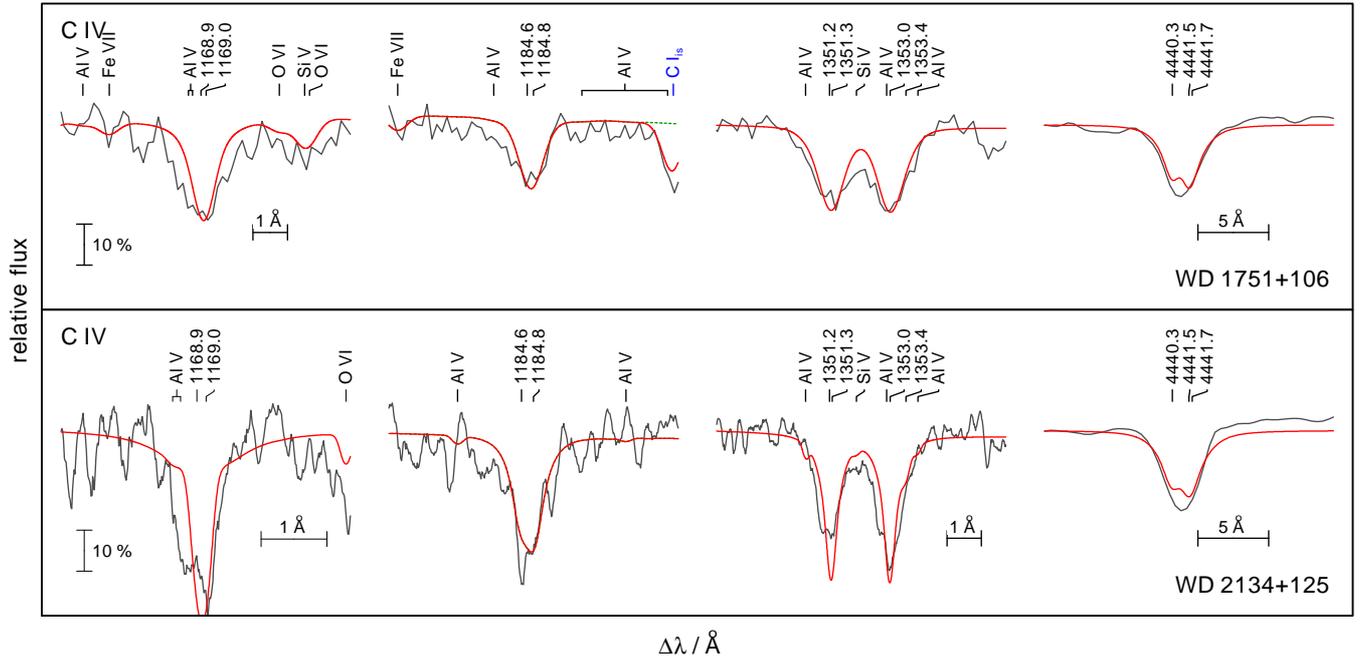}} 
   \caption{Synthetic spectra (red) around \ion{C}{iv} lines calculated from our final model compared with 
            observations (gray).}
   \label{fig:c}
\end{figure*}

\begin{figure*}
  \resizebox{\hsize}{!}{\includegraphics{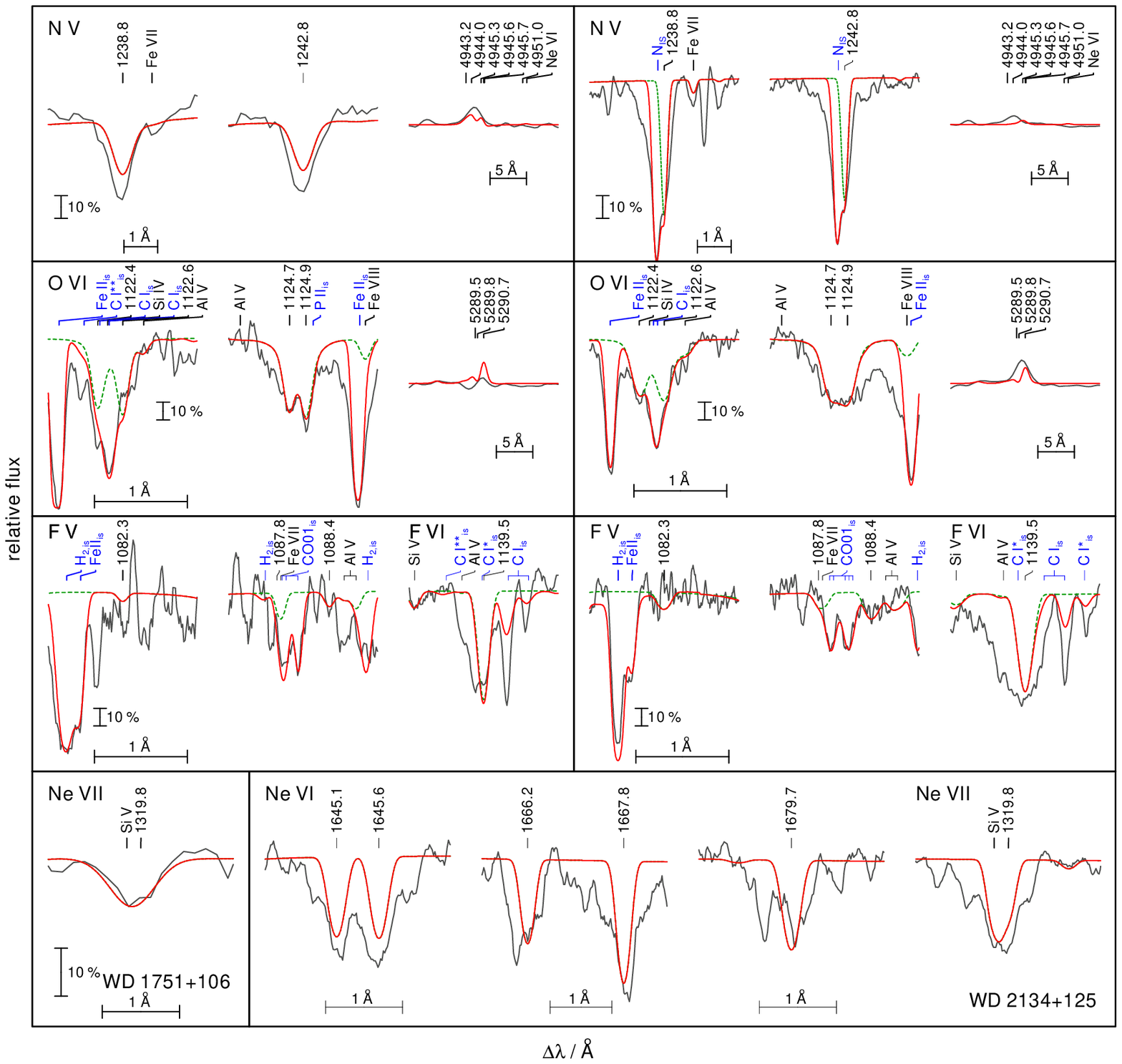}}
   \caption{Like Fig.\,\ref{fig:c}, for
             \ion{N}{v},
             \ion{O}{vi},
             \ion{F}{v-vi}, and
             \ion{Ne}{vi-vii}.
           }
   \label{fig:n-ne}
\end{figure*}

\begin{figure*}
  \resizebox{\hsize}{!}{\includegraphics{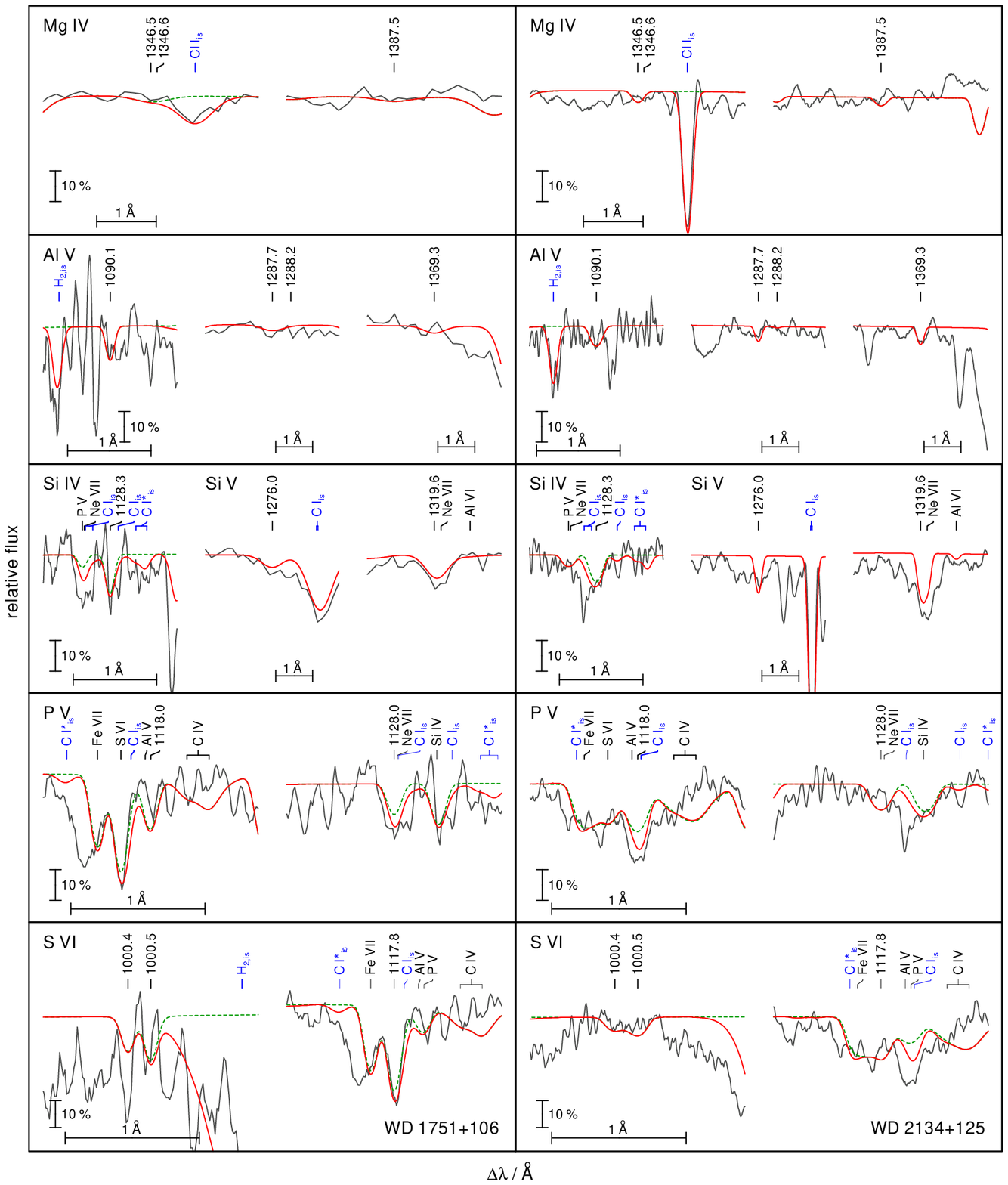}}
   \caption{Like Fig.\,\ref{fig:c}, for
           \ion{Mg}{iv}, 
           \ion{Al}{v}, 
           \ion{Si}{iv-v}, 
           \ion{P}{v}, and 
           \ion{S}{vi}.
           }
   \label{fig:mg-s}
\end{figure*}

\begin{figure}
  \resizebox{\hsize}{!}{\includegraphics{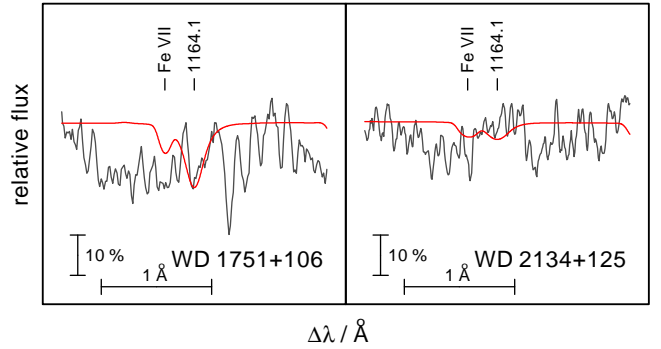}}
   \caption{Synthetic spectra around \Jonw{Ar}{viii}{1164.1} calculated from our final models
            compared with the FUSE observations.}
   \label{fig:ar}
\end{figure}

\begin{figure*}
  \resizebox{\hsize}{!}{\includegraphics{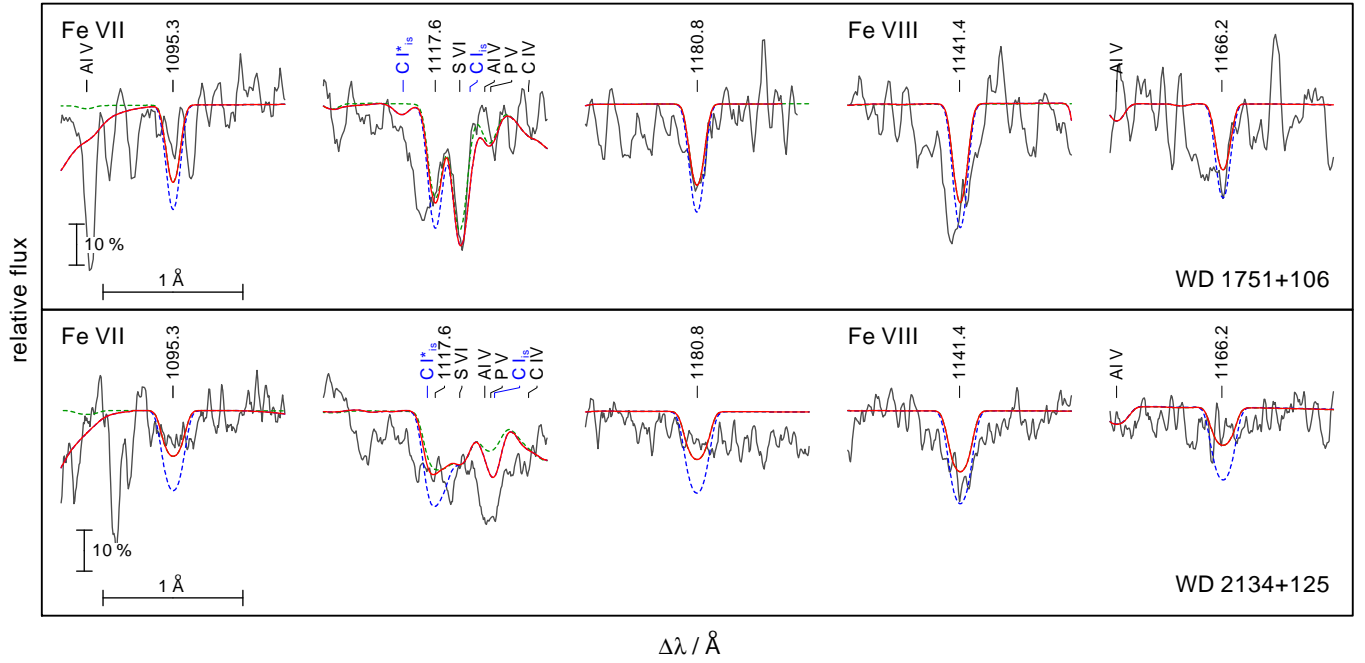}}
   \caption{Synthetic spectra around \ion{Fe}{vii} and \textsc{viii} lines calculated from our final models with $[\mathrm{Fe}]=\textcolor{black}{-0.4}$ (red, full) and solar (blue, dashed)
            compared with the FUSE observation.}
   \label{fig:fe}
\end{figure*}

In the following paragraphs, we discuss all elements, that were considered in this analysis. \textcolor{black}{To determine the abundances, we varied them in subsequent line formation calculations in steps of 0.2\,dex or smaller.} The abundances were derived by line-profile fits \textcolor{black}{and evaluation by eye}. For illustration, some representative spectral lines are shown in Figs.\,\ref{fig:c}-\ref{fig:fe}. These values are affected by typical errors estimated to 0.3\,dex \textcolor{black}{by redoing the abundance determination for models at the edges of the error range for \Teff and \logg (we used a model with \Teffw{120\,000}, \loggw{5.5} and one with \Teffw{110\,000} and \loggw{5.7})}. If no line identification was possible, we determined upper limits by reducing the abundance until the strongest computed lines become undetectable within the noise of the spectrum. The results are summarized in Table\,\ref{tab:finab}. The whole FUSE spectra compared with our final models are shown in Fig.\,\ref{fig:fuseall} and the GHRS spectrum of \wda as well as the STIS spectrum of \wdn in Fig.\,\ref{fig:stisall}. \textcolor{black}{The solar abundances are taken from \citet{asplundetal2009,scottetal2015a,scottetal2015b,grevesseetal2015}}.

\paragraph*{Carbon.} 
Detailed line-profile fits were performed for 
\Ionww{C}{4}{1168.9, 1169.0, 1184.6} in the FUSE observations,  
\Ionww{C}{4}{1351.2, 1351.3, 1353.0} in the GHRS or STIS observations, and 
the \ion{C}{iv} lines at $3685-3691$\,{\AA}, $4440-4442$\,{\AA}, $4785-4790$\,{\AA}, $5016-5019$\,{\AA}, and $6591-6593$\,{\AA}
in the UVES SPY observations \sA{fig:c}.
We achieve
$[\mathrm{C}]= \log\,\mathrm{(abundance/solar\ abundance)} =\textcolor{black}{2.0}$ for \wda and 
$[\mathrm{C}]=\textcolor{black}{2.1}$ for \wdn. 

\paragraph*{Nitrogen.}
The photospheric \Ionww{N}{5}{1238.8, 1242.8} resonance doublet in the STIS spectrum of \wdn is blended by strong
interstellar absorption. Therefore, we used 
\Ionww{N}{5}{4943.2, 4944.0, 4945.3 4945.6, 4945.7} 
in addition \sA{fig:n-ne} and determined
$[\mathrm{N}]=\textcolor{black}{0.6}$  for \wda and 
$[\mathrm{N}]=\textcolor{black}{-0.3}$ for \wdn. 
 
\paragraph*{Oxygen.}
The determination of the O abundance in \wdn is hampered by the fact that all useful lines in the 
UV and FUV are either blended with interstellar lines or display strong P\,Cygni profiles (e.g\@. 
\Ionw{O}{5}{1371.3} and \Ionww{O}{6}{1031.9, 1037.6}). We used \Ionww{O}{6}{1122.4, 1122.6, 1124.7, 1124.9} in the 
FUSE spectra and \Ionww{O}{6}{5289.5, 5289.8, 5290.7, 5292.0} in the UVES SPY spectra \sA{fig:n-ne} to determine $[\mathrm{O}]=\textcolor{black}{-0.2}$ 
for \wda and $[\mathrm{O}]=\textcolor{black}{-0.1}$ for \wdn. 
Furthermore, we employed PoWR to calculate wind profiles. 
Details on the wind models are given in Section \ref{sect:powr}. 

\paragraph*{Fluorine.}
The strong line \Ionw{F}{6}{1139.5} shows a P\,Cygni profile. For an abundance determination with our static models, we analysed \Ionww{F}{5}{1082.3, 1087.8, 1088.4} \sA{fig:n-ne}, which 
are reproduced best with an 
abundance of 
$[\mathrm{F}]=\textcolor{black}{1.0}$ for \wda and 
$[\mathrm{F}]=\textcolor{black}{1.5}$ for \wdn \citep[the same value was measured by][]{werneretal2005}. These abundances 
exceed the values of \citet{ringatetal2011} and \citet{ziegleretal2009a} but agree with the value of 
$[\mathrm{F}]=1.2$ \citep{reiffetal2008} \textcolor{black}{for \wdn}, who employed a wind model for their analysis.
We examined the profile of \Ionw{F}{6}{1139.5} in our PoWR wind model and verified the newly
determined abundances. 

\paragraph*{Neon.}
All lines of \Ion{Ne}{5} with observed wavelengths in the FUSE and STIS wavelength range that are available from the National Institute for Standards and Technology (NIST) Atomic Spectra Database\footnote{\url{https://physics.nist.gov/PhysRefData/ASD/lines_form.html}} are affected by an wavelength uncertainty of 1.5\,{\AA} or blended with interstellar absorption like \Ionw{Ne}{5}{946.9}.
\Ionww{Ne}{6}{1645.1, 1645.6, 1666.2, 1667.8, 1679.7} are visible in the STIS spectrum of \wdn and used for an estimate of the Ne abundance \sA{fig:n-ne}. \Ionw{Ne}{7}{1319.8} is also detectable but blended with \Ionw{Si}{5}{1319.6}. The optical lines Ne\,{\textsc{vii}}~$\lambda\lambda\,3643.6$, $3853.5$, $3866.7$, $3873.3$, $3894.0$, $3905.3$, $3912.0$\,{\AA} are very weak.
We determine $[\mathrm{Ne}]=\textcolor{black}{1.2}$ for \wdn and pose an upper limit of $[\mathrm{Ne}]<\textcolor{black}{1.5}$ for \wda 
owing to the resolution of the GHRS observation and the fact that the strong \Ion{Ne}{6} lines are not within the GHRS range. \Ionw{Ne}{7}{973.3} exhibits a strong P\,Cygni profile. The 
wind profile of this line in the PoWR model confirms the Ne abundance.

\paragraph*{Magnesium.}
No Mg line can be identified in the observations. Based on the computed lines of Mg\,{\textsc{iv}}~$\lambda\lambda\,1346.5$, $1346.6$, $1382.5$, $1385.7$, $1387.5$\,{\AA} \sA{fig:mg-s}, we find upper limits of $[\mathrm{Mg}]<\textcolor{black}{0.5}$ for \wdn and $[\mathrm{Mg}]<\textcolor{black}{0.8}$ for \wda. The latter value is higher due to the lower resolution of the GHRS observation compared to the one obtained with STIS.

\paragraph*{Aluminum.}
\Ionww{Al}{5}{1090.1, 1287.7, 1288.2, 1369.3} are the most prominent Al lines in the synthetic spectra \sA{fig:mg-s}. 
We determined an abundance of $[\mathrm{Al}]=\textcolor{black}{0.6}$ for \wdn
an upper limit of $[\mathrm{Al}]<\textcolor{black}{0.7}$ for \wda.

\paragraph*{Silicon.}
In the STIS and GHRS observations of both stars, the \Ionww{Si}{4}{1393.8, 1402.8} resonance doublet is blended by 
interstellar absorption lines.  
Thus, our abundance determination is based on \Ionw{Si}{4}{1128.3} and Si\,{\textsc{v}}~$\lambda\lambda\,1118.8$, $1251.4$, $1276.0$, $1291.4$, $1319.6$\,{\AA} \sA{fig:mg-s}. 
These lines indicate $[\mathrm{Si}]=\textcolor{black}{-0.6}$ for \wda.
For \wdn, we used the same lines as well as  \Ionww{Si}{5}{1465.5, 1582.7} and measured
$[\mathrm{Si}]=\textcolor{black}{-0.7}$. 

\paragraph*{Phosphorus.}
We used the strongest P lines, namely \Ionww{P}{5}{1118.0, 1128.0}, in the theoretical spectra to establish upper 
limits of $[\mathrm{P}]<\textcolor{black}{0.3}$ for \wda and $[\mathrm{P}]<\textcolor{black}{0.4}$ for \wdn \sA{fig:mg-s}.

\paragraph*{Sulfur.}
The most prominent S lines in the FUSE spectrum of \wda and \wdn, S\,{\textsc{vi}}~$\lambda\lambda\,933.4$, $944.5$\,{\AA} are both blended by interstellar $\mathrm{H}_2$ lines and thus it is uncertain to derive the S abundance. 
From S\,{\textsc{vi}}~$\lambda\lambda\,1000.4$, $1000.5$, $1117.8$\,{\AA} \sA{fig:mg-s}, we determine $[\mathrm{S}]=\textcolor{black}{-0.1}$ and 
$[\mathrm{S}]\le \textcolor{black}{-0.6}$ for \wda and \wdn, respectively. 

\paragraph*{Chlorine.}
\Ionww{Cl}{7}{949.0, 949.1, 996.7, 997.0} are present in the synthetic spectra but cannot be identified in 
the FUSE observations of both stars. Based on these lines, we determined upper limits of $[\mathrm{Cl}]<\textcolor{black}{1.1}$ 
and $[\mathrm{Cl}]<\textcolor{black}{1.0}$ for \wda and \wdn, respectively. 

\paragraph*{Argon.}
The strong \Ionw{Ar}{7}{1063.6} line is blended by interstellar absorption and thus can not be used to 
derive an
abundance value for argon. We used \Ionw{Ar}{8}{1164.1} to derive an upper limit of $[\mathrm{Ar}]<\textcolor{black}{0.3}$ 
for \wda and $[\mathrm{Ar}]<\textcolor{black}{-0.3}$ for \wdn (Fig\@.\ref{fig:ar}).

\paragraph*{Calcium.} 
The strongest line in the synthetic spectra, namely \Ionw{Ca}{9}{1116.0}, is blended with a strong 
interstellar $\mathrm{H}_2$ absorption feature. For \wda, this line appears in the wing 
of the $\mathrm{H}_2$ line and is used to derive an upper limit of $[\mathrm{Ca}]<\textcolor{black}{0.0}$.

\paragraph*{Chromium.} 
None of the \Ion{Cr}{7} lines appearing in the synthetic spectrum has been identified in the observation. 
We used \Ionww{Cr}{7}{1170.1, 1186.6, 1187.3} to determine upper limits of $[\mathrm{Cr}]<\textcolor{black}{2.0}$ for \wda and \wdn.  

\paragraph*{Iron.} 
In previous analysis of \pnn, \citet{miksaetal2002} found subsolar values of at least one dex for 
Fe. For the CS of \pna, they found a solar upper limit for Fe. With extended model atoms, \citet{loebling2018} checked and correct this result and 
determined $[\mathrm{Fe}]=\textcolor{black}{-0.4}$ for \wda and $[\mathrm{Fe}]=\textcolor{black}{-0.8}$ for \wdn. These values are supported also by our final model including all opacities of 31 elements (Fig\@. \ref{fig:fe}). 
Previous analyses that assumed a larger deficiency, did not take line broadening due to stellar rotation \textcolor{black}{or macro turbulence} into account explaining the lower Fe abundances. A solar Fe abundance can be ruled out since the computed lines of \Ion{Fe}{7} appear too strong \sA{fig:fe}

\paragraph*{Nickel.} 
The dominating ionization stages of the IGEs in the expected parameter 
regime are {\sc vii} and {\sc viii}. Due to the lack of POS lines of Ni in 
these stages in Kurucz's line list \citep{kurucz1991,kurucz2009}
for the spectral ranges of FUSE, STIS, and GHRS,
no Ni lines could be detected and identified. In their analysis, \citet{ziegleretal2009a} used
\Ion{Ni}{6} lines and determined upper Ni abundance limits only. Their subsolar values may again be a result of not taking \textcolor{black}{additional broadening due to} rotation \textcolor{black}{or macro turbulence} into account. Furthermore, they assumed lower temperatures and higher gravities for both stars. The \Ion{Ni}{6} lines are very sensitive to \Teff and are significantly stronger for a model with \Teff reduced by 10\,kK \sA{fig:ni2}.\\\\
No strong lines in the spectra of both stars were found from the elements Sc, Ti, V, Mn, Ni, and Co. 
Therefore, these elements were combined to a generic model atom \citep{rauchdeetjen2003}.
All IGEs were taken into account with solar abundances ratios normalized to the Fe abundance 
in the final model calculations.

\paragraph*{Zinc.} 
The strong lines Zn\,{\scriptsize\Ionst{5}}~$\lambda\lambda\,1132.3, 1123.7, 1133.0, 1133.1,$ $1133.3, 1133.5, 1174.3, 1180.0$\,{\AA} 
appear in the synthetic spectra but could not be identified in the observations. Thus, we can only derive an upper limit of 
$[\mathrm{Zn}]<\textcolor{black}{1.0}$ for both stars. 

\paragraph*{Gallium.} 
We used the strongest line \Ionw{Ga}{6}{1006.9} to determine an upper limit of $[\mathrm{Ga}]<\textcolor{black}{2.0}$ for \wda and \wdn.

\paragraph*{Germanium.} 
\Ionww{Ge}{6}{920.5, 926.8, 988.2} are partly blended by interstellar absorption features. The analysis yields an upper 
limit of $[\mathrm{Ge}]<\textcolor{black}{2.0}$ for both stars.

\paragraph*{Krypton.} 
\Ionww{Kr}{7}{1166.9, 1169.6, 1195.6, 1284.6} are present in the models. We used these lines to 
measure an upper limit of $[\mathrm{Kr}]<\textcolor{black}{3.5}$ for both stars.

\paragraph*{Zirconium.} 
By analyzing STIS observation of \wdn around the computed lines Zr\,\textsc{vii}~$\lambda\lambda\,1233.6$, $1235.0$, $1376.6$\,{\AA} ,we derived an 
upper limit of $[\mathrm{Zr}]<\textcolor{black}{3.0}$. Due to the fact that these lines are located in the range of the lower-resolution 
GHRS observation of \wda, we were not able to derive a reasonable value for this star.

\paragraph*{Tellurium.}
We used \Ionw{Te}{6}{1071.4}, the strongest computed line in the FUSE range, to ascertain $[\mathrm{Te}] \le \textcolor{black}{4.0}$
for \wda and $[\mathrm{Te}] \le \textcolor{black}{3.5}$ for \wdn.

\paragraph*{Iodine.}
The strongest computed lines I\,\textsc{vi}~$\lambda\lambda\,911.2$, $915.4$, $919.2$\,{\AA} are useless for the abundance measurement, since they
are blended by interstellar absorption. Based on \Ionww{I}{6}{1045.4, 1120.3,1153.3}, we derived upper limits of 
$[\mathrm{I}]<\textcolor{black}{4.6}$ and $[\mathrm{I}]<\textcolor{black}{5.0}$ for \wda and \wdn, respectively.

\paragraph*{Xenon.} 
We used the strongest line \Ionw{Xe}{7}{995.5} to determine an upper limit of $[\mathrm{Xe}]<\textcolor{black}{4.0}$ for \wdn. 
The quality of the FUSE observation of \wda around this line does not suffice to determine the Xe abundance.

\paragraph*{Selenium, strontium, molybdenum, and barium.} 
Even if we increase the Se, Sr, Mo, and Ba abundances in the synthetic models to thousand times solar, no lines of these 
elements appear in the computed spectra. \\\\
In our final models, all TIEs are taken into account with solar abundance ratios normalized to the determined Fe 
abundance value. The temperature and density structure and the ionization fractions of all ions considered in the final model for \wdn are shown in \ab{fig:ionfrac}.

\begin{figure*}
  \resizebox{\hsize}{!}{\includegraphics{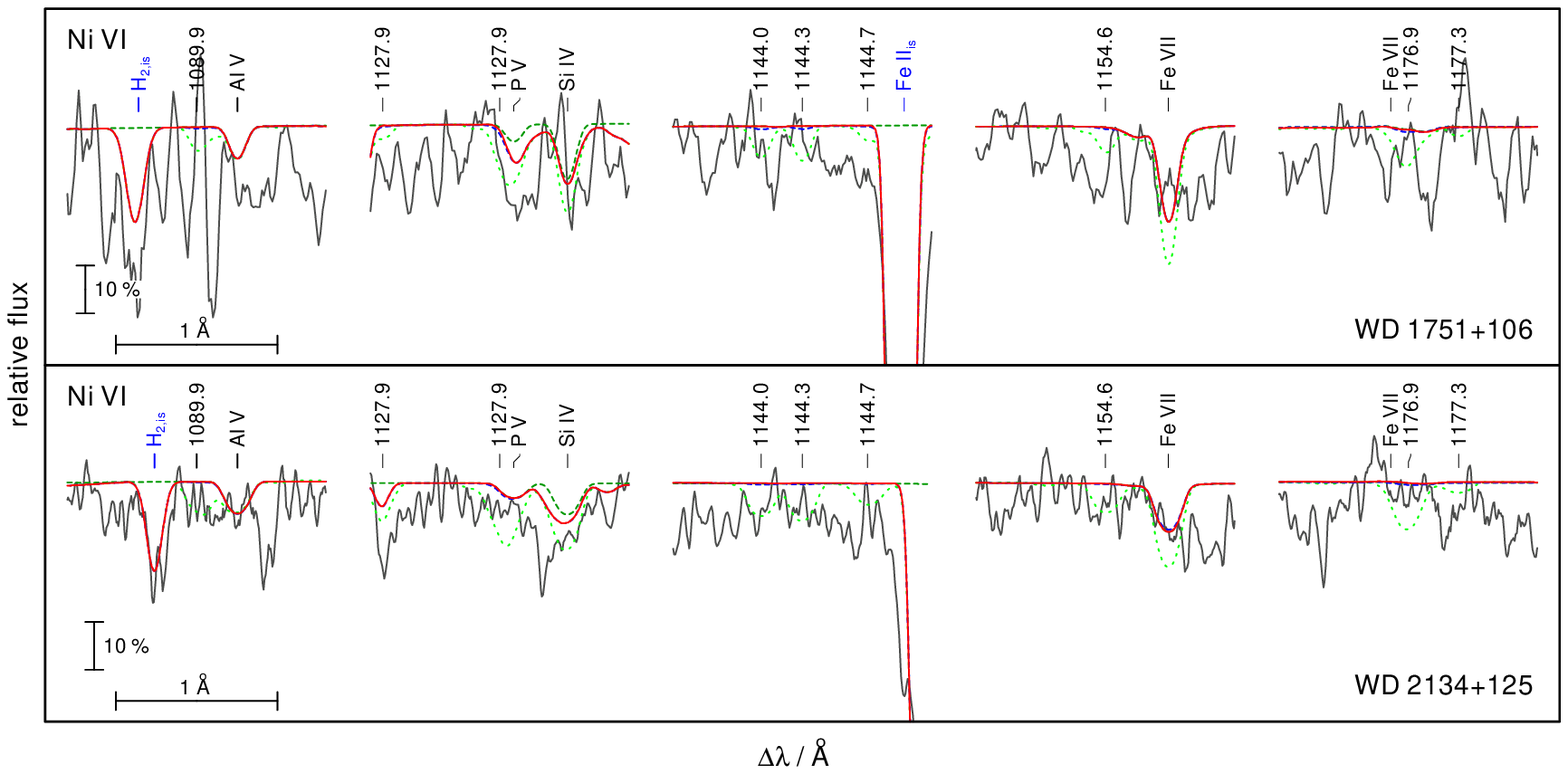}}
   \caption{Top panel: Synthetic spectra around \ion{Ni}{vi} lines calculated from our final models for \wda with $\Teff = 115$\,kK and $[\mathrm{Ni}]=\textcolor{black}{-0.4}$ (red, full) and one dex supersolar (blue, dashed) and $\Teff = 105$\,kK and Ni one dex supersolar (green, dotted). Bottom: For \wdn with $\Teff = 115$\,kK and  $[\mathrm{Ni}]=\textcolor{black}{-0.8}$ (red, full) and one dex supersolar (blue, dashed) and $\Teff = 105$\,kK and Ni one dex supersolar (green, dotted).}
   \label{fig:ni2}
\end{figure*}

\section{Stellar wind and mass loss}
\label{sect:powr}
At \Teffw{115\,000} and \loggw{5.6} the stars have luminosities of almost 
4\,000\,$L_\odot$ \sK{sect:mld}.
They are located close to the Eddington limit and experience mass loss due to a
radiation-driven wind \citep[cf\@.][]{pauldrachetal1988} and, hence, exhibit prominent P\,Cygni profiles in their UV
spectra \sA{fig:wind}.
\citet{koesterkewerner1998} and \citet{koesterkeetal1998} investigated the wind properties
of \wdn\ by means of NLTE models for spherically expanding atmospheres
and determined the mass-loss rate $\dot{M}$ 
and the terminal wind velocity $v_\infty$ from HST/GHRS and 
ORFEUS-SPAS\,II\footnote{Orbiting and Retrievable Far and Extreme Ultraviolet Spectrometer -- Space Pallet Satellite II}
observations, respectively.
They found 
$\log [\dot{M}\,/\,(M_\odot/\mathrm{yr})]=-7.3$ from \Ionww{C}{4}{1548.20, 1550.77}
and        
$\log [\dot{M}\,/\,(M_\odot/\mathrm{yr})]=-7.7$ from \Ionww{O}{6}{1031.91, 1037.61} 
and $v_\infty=3\,500\,\mathrm{km/s}$
 which is slightly lower than the former value of $v_\infty=3\,900\,\mathrm{km/s}$
of \citet{kaleretal1985} based on the analysis of spectra obtained with IUE.
\citet{guerreroetal2013} used lines of \Ion{O}{6} and found $3\,610\,\mathrm{km/s}$ for \wdn and $3\,000-3\,600\,\mathrm{km/s}$
 from the analysis of \Ion{O}{6} and \Ion{Ne}{7} lines for \wda.\\
For the analysis of the wind lines in the UV range we used the 
PoWR model atmosphere code. This is a  state-of-the-art NLTE code that accounts for mass-
loss, line-blanketing, and wind clumping. It can be employed for
a wide range of hot stars at arbitrary metallicities 
\citep[e.g.][]{hainich2014, hainich2015, reindl2014, oskinova2011, 
shenar2015,reindletal2017}, since
the hydrostatic and wind regimes of the atmosphere are treated
consistently \citep{sander2015}. 
The NLTE radiative transfer
is calculated in the co-moving frame. 
Any model can be specified by its luminosity $L$, stellar temperature
$T_\mathrm{eff}$, surface gravity
$g$, and mass-loss rate $\dot{M}$ as main parameters.
In the subsonic region, the velocity field is defined such that
a hydrostatic density stratification is approached 
\citep{sander2015}. 
In the supersonic region, the wind velocity field $v(r)$ is
pre-specified assuming the so-called $\beta$-law 
\citep{castor1975}. Wind clumping is taken into account in first-order
approximation \citep{hamann2004} with a density contrast 
$D = \varrho_\text{cl}/ \langle \varrho \rangle$ between the clumps and a smooth wind of same mass-loss rate. 
As we do not assume an interclump medium, $D=f_V^{-1}$.\\
We adopted the stellar parameters from the TMAP analysis  as given in
\ta{tab:finab}. Our calculations include complex model atoms for H, He, C,
N, O, F, Ne, Si, P, S, and the iron group elements Sc, Ti, V, Cr, Mn,
Fe, Co, Ni. \\
The only P Cygni line that is not saturated and is
sensitive enough to the mass-loss rate is \Ionww{C}{4}{1548,1551}. For \wda the quality of the 
UV observation in this wavelength range is very poor. However, we found 
that the mass-loss rate must be 
$\log [\dot{M}\,/\,(M_\odot/\mathrm{yr})] \lesssim -8.1$ to
obtain a model that is compatible with the observation.
The STIS spectrum of \wdn in this wavelength range has a much
better quality. We obtained the best fit to the complicated line profile of
\Ionww{C}{4}{1548,1551} by models with a mass-loss rate of
$\log [\dot{M}\,/\,(M_\odot/\mathrm{yr})] \approx -8.1$. In both cases we assumed a
density contrast of $D=10$, which is typically found for H-deficient
CSPNe winds \citep{todt2008}.\\ 
The blue edges of the P Cygni profiles of 
\Ionww{O}{6}{1032,1038} and \Ionww{C}{4}{1548,1551}
were used to estimate the terminal wind velocity for \wdn of about 
$v_\infty = 3300 \pm 100\,$km/s and a $\beta=0.6$. 
Additional broadening due to depth dependent microturbulence with 
$v_\text{D} = 20\,$km/s in the 
photosphere, estimated from, e.g., the F\,\textsc{vi}$\,\lambda 1140$ line,
up to $v_\text{D} = 230\,$km/s in the outer wind
was taken into account and allows to fit the width of the O\,\textsc{vi} \sA{fig:wind} and
the C\,\textsc{iv}  resonance lines simultaneously.
Similar values have been obtained for \wda, i.e.\  
$v_\infty = 3500 \pm 100\,$km/s and $v_\text{D} = 50\,$km/s in the photosphere 
up to $v_\text{D} = 180\,$km/s in the outer wind.
At these high mass-loss rates, the stellar wind is coupled and the photosphere is chemically homogeneous \citep{unglaub2007,unglaub2008}.

\begin{figure*}
  \resizebox{\hsize}{!}{\includegraphics{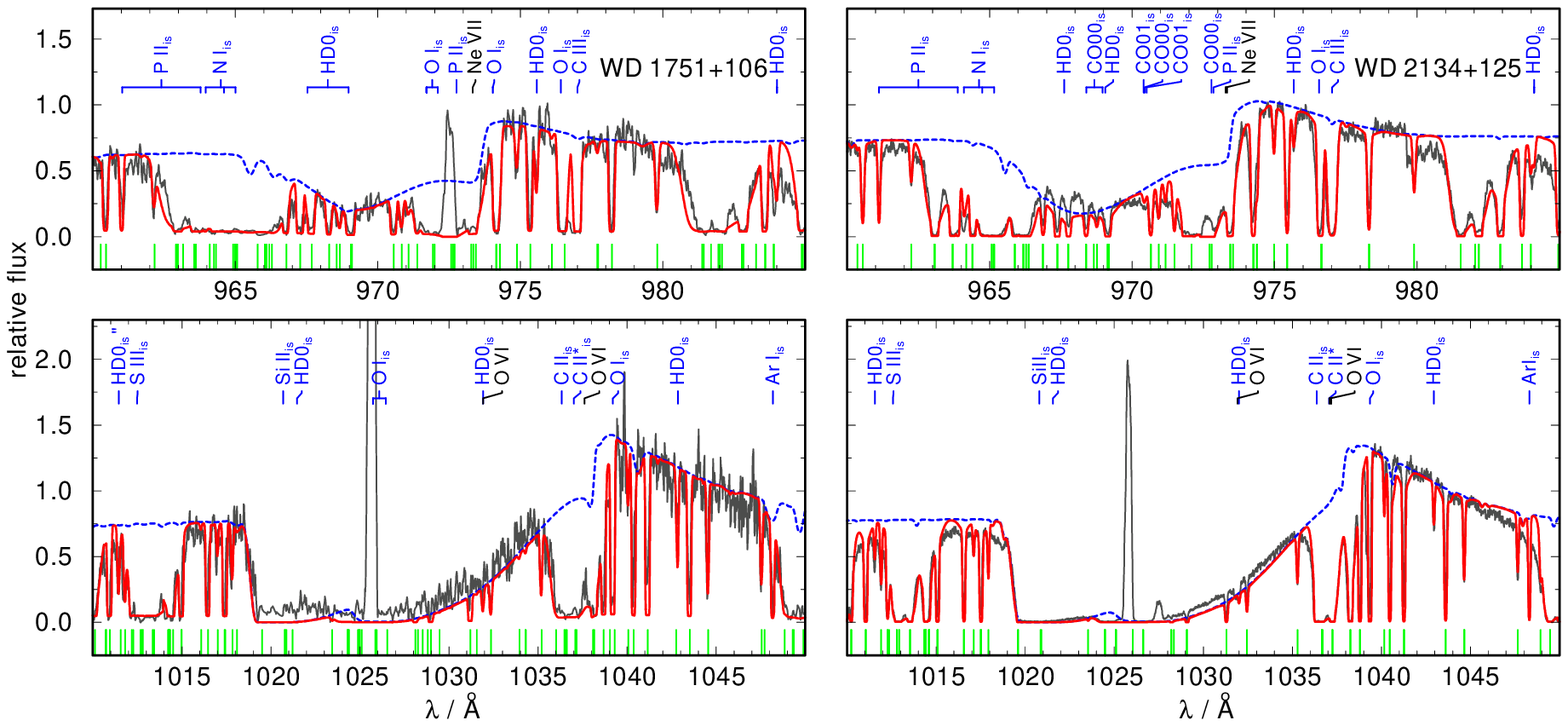}}
   \caption{Comparison of synthetic spectra calculated with
            PoWR (blue line, dotted) compared with the FUSE observation of \wda (left) and \wdn (right). In red a
            combined wind+ISM spectrum is shown. The wind models are calculated with a mass-loss rate of $\log [\dot{M}\,/\,(M_\odot/\mathrm{yr})] =-8.1$, and $v_\infty=3\,500\,\mathrm{km/s}$ for \wda and $v_\infty=3\,300\,\mathrm{km/s}$ for \wdn. 
            Upper panel: around \Jonw{Ne}{vii}{973.33}, lower: around \Jonww{O}{vi}{1031.91, 1037.61}.
            The green marks at the bottom of each panel indicate wavelengths of strong interstellar
            H$_2$ lines.
           }
   \label{fig:wind}
\end{figure*}

\section{Mass, luminosity, and distance}
\label{sect:mld}
The determination of the mass of \wda and \wdn is difficult since their evolutionary history
is not unambiguous. If they are AFTP stars no appropriate set of evolutionary tracks is available in the literature
to compare with. \citet{lawlormacdonald2006} presented a variety of calculations for the 
evolution of H-deficient post-AGB stars. The so-called AGB departure type V scenario
(departure from the AGB during a He flash) and the type IV scenario appear to be in the transition between AFTP 
and LTP \sT{tab:lawlor}. 
A different H/He ratio should have an influence on the result \citep[cf.][]{millerbertolamialthaus2007}. {To decide which grid of post-AGB tracks should be used, we calculated some AFTP models with {\tt LPCODE} \citep{althausetal2003, althausetal2005} \sT{tab:lawlor}. This was done by recomputing the end of the AGB evolution of three models presented in  \cite{millerbertolami2016} and tuning the mass loss at the end of the AGB phase as to enforce an AFTP event. These sequences have $(M_{\rm ZAMS}$, $Z_{\rm ZAMS}$, $M_{\rm f})=$ $(1.25M_\odot$, $0.01$, $0.566M_\odot)$, $(1.00M_\odot$, $0.001$, $0.550M_\odot)$, $(1.50M_\odot$, $0.001$, $0.594M_\odot)$ \sT{tab:lawlor}}. At the location of the stars in the \logg-\Teff diagram, the tracks for VLTP and AFTP stars coincide \sA{fig:logglogteff}. Thus, the approach of using VLTP tracks for the determination of the mass is acceptable.\\
We find $M=0.57^{+0.07}_{-0.04}\,M_\odot$ for \wda and \wdn. Using the initial-final mass relation of \citet{cummingsetal2018}, these stars originate from progenitors with initial mass of about $1.0$ to $1.1\,M_\odot$.
From the 0.515, 0,530, 0,542, 0,565, 0.584, 0.609, and 0.664$\,M_\odot$ tracks \sA{fig:logglogteff}, we determine the luminosity of $\log L\, / \, L_\odot = 3.77 ^{+0.23}_{-0.24}$ for \wda and \wdn.\\
The spectroscopic distances are calculated following the flux calibration\footnote{\url{http://astro.uni-tuebingen.de/~rauch/SpectroscopicDistanceDetermination.gif}} of \citet{heberetal1984}, 
\begin{equation}
\label{Eq:fv}
f_\mathrm{V} = 3.58\times 10^{-9}\times 10^{{\rm -0.4}m_\mathrm{V_0}}\, \mathrm{erg\,cm^{-2}\,s^{-1}}\,{\text{\normalfont\AA}}^{-1}
\end{equation}
\noindent
with 
$m_\mathrm{V_0} = m_\mathrm{V} - 2.175 c$, 
$c = 1.47 E_\mathrm{B-V}$. 
We take $m_\mathrm{V} = 14.75\pm 0.13$ \citep{ackeretal1992} and $c = 0.390 \pm 0.015$
using our determination of $E_\mathrm{B-V}$ for \wda and $m_\mathrm{V} = 13.68\pm 0.25$ 
\citep{ackeretal1992} and $c = 0.199 \pm 0.015$ for \wdn.
The distance is derived from
\begin{equation}
\label{Eq:d}
d\,/\,pc = 7.11\times 10^{4} \sqrt{H_\nu (M/M_\odot)\times 10^{(0.4m_\mathrm{V_0}-\log g)}}\,.
\end{equation}
\noindent
The Eddington flux at $\lambda_\mathrm{eff} = 5454\,{\text{\normalfont\AA}}$ of our final model atmospheres including all 31 elements is
$H_\nu = (1.89\pm 0.19) \times 10^{-3} \mathrm{erg\,cm^{-2}\,s^{-1}\,Hz^{-1}}$ for \wda and $H_\nu = (1.88\pm 0.20) \times 10^{-3} \mathrm{erg\,cm^{-2}\,s^{-1}\,Hz^{-1}}$ for \wdn.
We derive distances of 
$d= 2.23^{+0.31}_{-0.33}\,\mathrm{kpc}$ for \wda and 
$d= 1.65^{+0.32}_{-0.31}\,\mathrm{kpc}$ for \wdn. 
\wda is located $0.67^{+0.10}_{-0.10}\,\mathrm{kpc}$ above the Galactic plane and \wdn has a depth below the 
Galactic plane of $0.78^{+0.15}_{-0.15}\,\mathrm{kpc}$. 
Taking into account the angular sizes of the nebulae measured from narrow-band images by \citet{rauch1999},
the nebula shells of \pna and \pnn have radii of 
$R= 0.48^{+0.07}_{-0.07}\,\mathrm{pc}$ and 
$R= 0.42^{+0.08}_{-0.08}\,\mathrm{pc}$, respectively.
With the measured expansion velocity of 
$40 \pm 2\,\mathrm{km/s}$ for \pna and 
$38 \pm 2\,\mathrm{km/s}$ for \pnn \citep{pereyraetal2013}, the expansion times are
$11\,600^{+1\,600}_{-1\,700}\,\mathrm{yrs}$ and 
$10\,900^{+2\,100}_{-2\,000}\,\mathrm{yrs}$, respectively. These dynamical timescales place a lower limit to the actual age of the PNe since velocity gradients and the acceleration over time are not taken into account \citep{gesickietal2000}. \citet{dopitaetal1996} derived typical correction factors of 1.5. 
Both stars are targets of the Gaia mission and contained in the data made public in the second data release (DR2). 
Gaia measured parallaxes of $0\farcs 431 \pm 0\farcs061$ and $0\farcs615 \pm 0\farcs059$ \citep{gaiavizier2018} for \wda (ID\,4488953930631143168) and \wdn (ID\,1770058865674512896), respectively. This corresponds to relative errors of $14.1\,\%$ and $9.5\,\%$. From these values, \citet{baillerjonesetal2018} derived distances of $2.19^{+0.35}_{-0.27}\,$\,kpc for \wda and $1.55 ^{+0.16}_{-0.13}$\,kpc for \wdn. These values are in good agreement with our spectroscopic distance determination and validate the mass determination by spectroscopic means using stellar atmosphere models and evolutionary tracks.

\begin{table*}
\setlength{\tabcolsep}{0.50em}
\caption{Photospheric abundances of \wda and \wdn compared with evolutionary models.}
\label{tab:lawlor}
\begin{tabular}{rcr@{.}lr@{.}lr@{.}lr@{.}lr@{.}lr@{.}lp{7cm}}
\hline
\hline
\noalign{\smallskip}
\multicolumn{1}{c}{\Teff} & & \multicolumn{2}{c}{H} & \multicolumn{2}{c}{He} & \multicolumn{2}{c}{C} & \multicolumn{2}{c}{N} & \multicolumn{2}{c}{O} & \multicolumn{2}{c}{Ne}\\ 
\cline{3-14}
\multicolumn{15}{c}{} \vspace{-5.5mm}\\ 
 & $\log (g\,/\,\textrm{cm}/\textrm{s}^2)$& \multicolumn{12}{c}{} & comment \vspace{-7mm}\\ 
  \multicolumn{1}{c}{(K)}  &    & \multicolumn{12}{c}{(mass fraction)} \\
\hline
\noalign{\smallskip}
115\,000 & 5.6 & 0&15 & 0&52 & 0&31 & 0&0003  & 0&0033 & 0&0019& \wdn, our atmosphere model \\
115\,000 & 5.6 & 0&25 & 0&46 & 0&27 & 0&0026  & 0&0044 & 0&012& \wda, our atmosphere model \\
 84\,000 & 5.0 & 0&444 & 0&539 & 0&012 & 0&002  & 0&002 & 0&0008 &\citet[][AGB departure type IV]{lawlormacdonald2006} \\
140\,000 & 6.0 & 0&106 & 0&794 & 0&085 & 0&002  & 0&012 & 0&003 &\citet[][AGB departure type IV]{lawlormacdonald2006}\\
 87\,000 & 5.0 & 0&565 & 0&427 & 0&005 & 0&0009  & 0&0009 & 0&0004 &\citet[][AGB departure type V]{lawlormacdonald2006} \\
150\,000 & 6.0 & 0&164 & 0&775 & 0&051 & 0&0014  & 0&0057 & 0&002 &\citet[][AGB departure type V]{lawlormacdonald2006}\\
  &  & 0&197 & 0&450 & 0&296 & 0&0001  & 0&056 & 0&00078 & AFTP model $M_\mathrm{f}=0.550$, $Z=0.001$ \\
  &  & 0&137 & 0&390 & 0&357 & 0&0006  & 0&104 & 0&0085 & AFTP model $M_\mathrm{f}=0.566$, $Z=0.01$ \\
  &  & 0&219 & 0&445 & 0&258 & 0&0007  & 0&055 & 0&021 & AFTP model $M_\mathrm{f}=0.594$, $Z=0.001$ \\
\hline
\end{tabular}
\end{table*}

\begin{figure}
  \resizebox{\hsize}{!}{\includegraphics{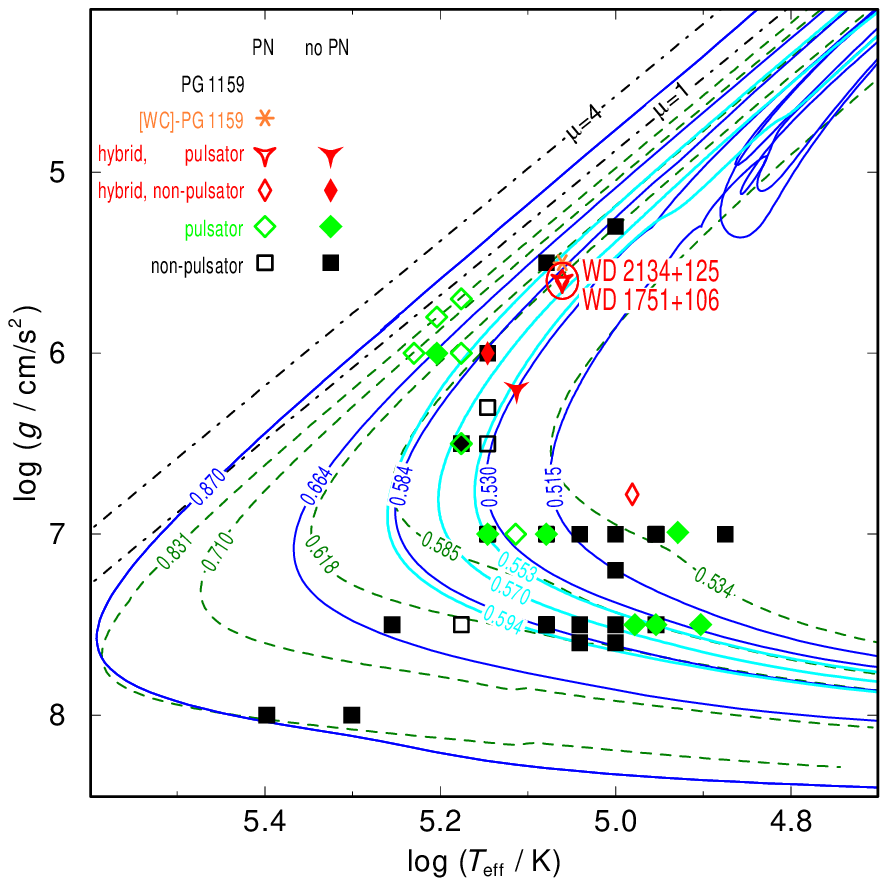}}
   \caption{Positions of \wda (\pna) and \wdn (\pnn) with their error ellipses 
            and related objects in the $\log \Teff - \logg$ plane
            compared with evolutionary tracks (labeled with the respective masses in M$_\odot$) of 
            VLTP stars \citep[][blue full lines]{millerbertolamialthaus2006}, of
            H-burning post-AGB stars
            \citep[calculated with initial solar metallicity,][green dashed lines]{millerbertolami2016}, and of AFTP stars (cyan, thick lines).\textcolor{black}{The dashed-dotted $\mu=1$ and $\mu=4$ lines indicate the Eddington limits for pure H and He atmospheres, respectively.}}
   \label{fig:logglogteff}
\end{figure}

\begin{figure*}
  \resizebox{\hsize}{!}{\includegraphics{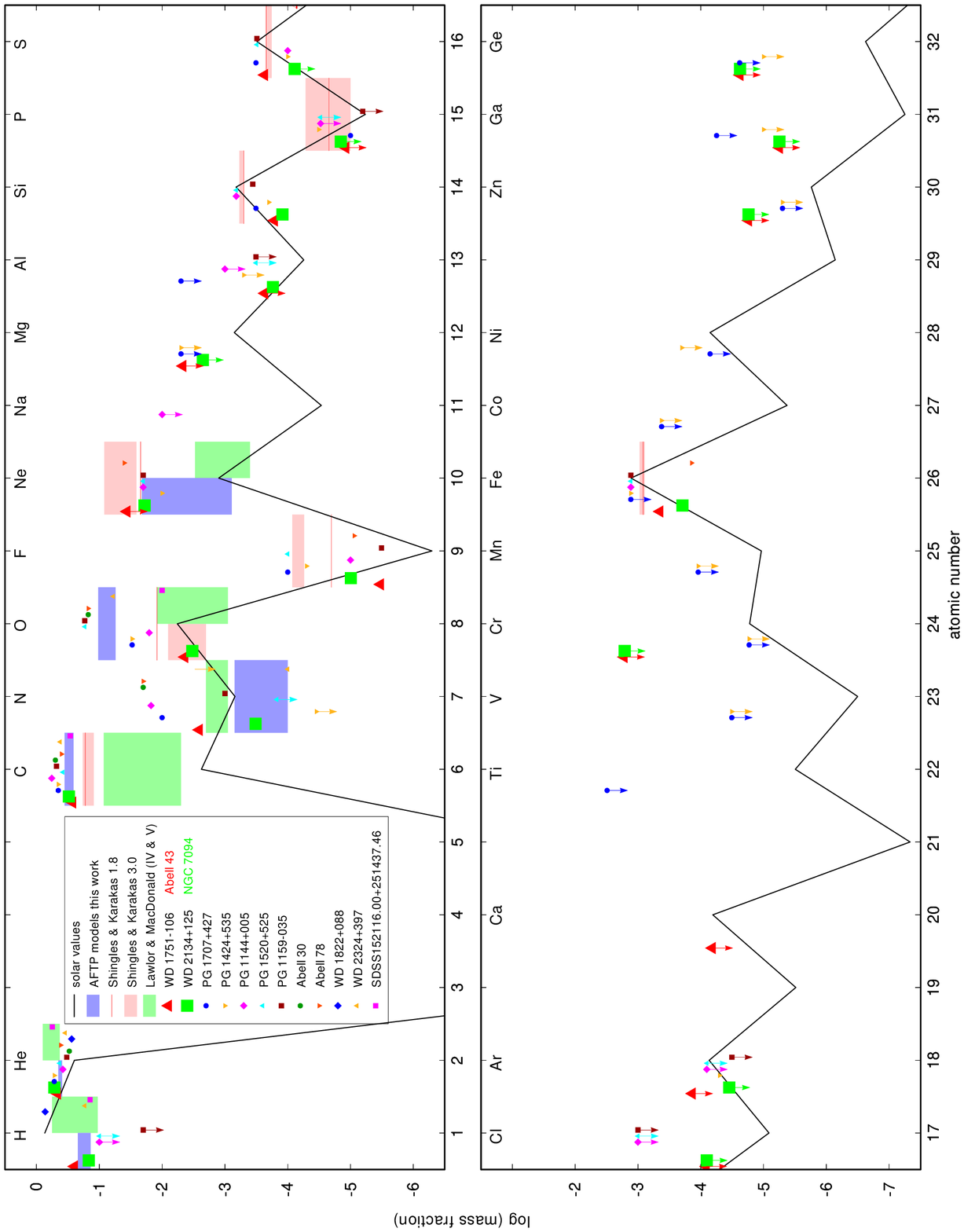}}
   \caption{Photospheric abundances for a set of PG\,1195 stars and calculated ranges from evolutionary models. For the models of \citet{shingles2013}, the initial mass of the star in $M_\odot$ is given in the legend box. 
            Upper limits are indicated with arrows.
            The solid black lines indicates solar abundances.
            }
   \label{fig:mfall}
\end{figure*}
  
\section{Discussion}
\label{sect:results}
Our aim was to determine the element abundances of \wda and \wdn beyond He and H. The results are shown in \ta{tab:finab}. By comparing our results to the abundances of other 
post-AGB stars and evolutionary models, we are able to conclude constraints for nucleosynthesis processes and evolutionary channels.
The H-deficient nature of \wda and \wdn suggests that here, as in other H-deficient stars, we see 
nuclear processed material on the surface that has formed either during the progenitor 
AGB phase or the same mixing and burning processes that lead to the H-deficiency in the 
first place. Figure\,\ref{fig:mfall} illustrates the following sections.

\subsection{Comparison to other hybrid PG\,1159 stars}
\label{sect:comphyb}
The group of known hybrid PG\,1159 stars comprises the CS of \object{Sh\,2$-$68} and \object{HS\,2324+3944} 
\citep[WD\,1822+008 and WD\,2324+397, respectively, ][]{mccooksion1999, mccooksion1999cat}
and \object{SDSS\,152116.00+251437.46} \citep{wernerherwig2006,werneretal2014},
besides the two program stars of this work. \textcolor{black}{The known atmospheric parameters for these objects are summarized in \ta{tab:comp}.} For \object{WD\,1822+008}, \citet{gianninasetal2010} found 
$T_\mathrm{eff} = 84\,460$\,K and $\log g = 7.24$. Its position in the 
$\log T_\mathrm{eff}$-$\log g$ diagram \sA{fig:logglogteff} suggests that the star is already located 
close to the beginning of the WD cooling track and is thus further evolved than the two stars of this work.
The large distance of $1000 \pm 400$\,pc \citep{binnendijk1952} was reduced and better constrained to $399.7^{+11.8}_{-12.5}$\,pc \citep{baillerjonesetal2018} using Gaia data. With its a diameter of $400'' \pm 70''$
\citep{fesen1983}, the PN has a radius of $0.388^{+0.082}_{-0.078}$\,pc. Considering an expansion velocity of 
$7.5$\,km/s \citep{hippelein1990}, this yields a dynamic timescale of $50\,600 ^{+10\,700}_{-10\,200}$\,yrs 
which confirms the suggestion from the location in the $\log T_\mathrm{eff}$-$\log g$ diagram and 
assigns Sh\,2$-$68 to the group of oldest and largest PNe. It has a lower element ratio of
He/H compared to \wda and \wdn. This might result from ongoing depletion of heavier elements
from the atmosphere due to gravitational settling.\\
\object{WD\,2324+397} 
\citep[$T_\mathrm{eff}= 130\,000 \pm 10\,000$\,K and $\log g = 6.2 \pm 0.2$, ][]{dreizler1996} 
and
\object{SDSS\,152116.00+251437.46} 
\citep[$T_\mathrm{eff}= 140\,000 \pm 15\,000$\,K and $\log g = 6.0 \pm 0.3$, ][]{werneretal2014}
are members of this group without an ambient PN \citep{werner1997}.
\textcolor{black}{The C abundance is similar to the values determined for our two program stars. Considering the C/H ratio, \textcolor{black}{the value for \object{SDSS\,152116.00+251437.46} is close to the one of \wdn while the one for \object{WD\,2324+397} is slightly higher. Both their O/H ratios exceeds the values of our two program stars by a factor of two or even more.}}
The N/H ratio of the stars for which it is known is very low.  
The He/H ratio of \object{WD\,2324+397} resembles the value of 
\wda 
whereas \wdn has a higher He content 
similar to \object{SDSS\,152116.00+251437.46}.

\begin{table*}
\setlength{\tabcolsep}{0.38em}
\caption{Photospheric abundances of \wda and \wdn compared with other hybrid PG\,1159 stars and the CSPNe Abell\,30 and Abell\,78.}
\label{tab:comp}
\color{black}
\begin{tabular}{lrr@{.}lr@{.}lr@{.}lr@{.}lr@{.}lr@{.}lp{6.2cm}}
\hline
\hline
\noalign{\smallskip}
 & \multicolumn{1}{c}{\Teff} & \multicolumn{2}{c}{} & \multicolumn{2}{c}{H} & \multicolumn{2}{c}{He} & \multicolumn{2}{c}{C} & \multicolumn{2}{c}{N} & \multicolumn{2}{c}{O} \\ 
\cline{5-14}
\multicolumn{15}{c}{} \vspace{-5.2mm}\\ 
Object &  & \multicolumn{2}{c}{$\log (g\,/\,\textrm{cm}/\textrm{s}^2)$}& \multicolumn{10}{c}{} & Reference \vspace{-2.5mm}\\ 
&   \multicolumn{1}{c}{(K)}  &  \multicolumn{2}{c}{}  & \multicolumn{10}{c}{(mass fraction)} \\
\hline
\noalign{\smallskip}
\wdn                      & 115\,000  & \quad\quad 5 & 6  & 0&15  & 0&52  & 0&31  & 0&0003    & 0&0032 &  this work\\
\wda                      & 115\,000  & 5 & 6  & 0&25  & 0&46  & 0&27  & 0&0026    & 0&0044 &  this work \\
WD\,1822$+$008            &  84\,460  & 7 & 24 & 0&66  & 0&34  &\multicolumn{2}{c}{}&\multicolumn{2}{c}{}&\multicolumn{2}{c}{}&  \citet{gianninasetal2010}\\
WD\,2324$+$397            & 130\,000  & 6 & 2  & 0&\color{black}17  & 0&\color{black}35  & 0&\color{black}42  & 0&\color{black}0001    & 0&\color{black}06 &  \citet{dreizler1996}; \citet{dreizler1999}\\
\color{black}
SDSS\,152116.00+251437.46 & 140\,000  & 6 & 0  & 0&\textcolor{black}{14}  & 0&\textcolor{black}{56}  & 0&\textcolor{black}{29} &\multicolumn{2}{c}{}& 0&01  &\citet{werneretal2014}\\
Abell\,30                 & 115\,000  & 5 & 5  &\multicolumn{2}{c}{}& 0&41 & 0&40& 0&04 & 0&15 & \citet{leuenhagen1993} \\
Abell\,78                 & 117\,000  & 5 & 5  &\multicolumn{2}{c}{}& 0&30 & 0&50& 0&02 & 0&15 & \citet{toala2015}; \citet{werner1992} \\
\hline
\end{tabular}
\end{table*}

\subsection{Comparison to \object{Abell\,30} and \object{Abell\,78}}
\label{subsect:A3078}
Comparing the results of our analysis to the parameters known for the CSs of the PNe \object{Abell\,30} and \object{Abell\,78}
is of special interest, because these objects are located at almost the same position in the 
$\log T_\mathrm{eff}$-$\log g$ diagram. \textcolor{black}{Both are [WC]-PG\,1159 transition objects with 
$T_\mathrm{eff} = 115\,000$\,K and $\log g = 5.5$ \citep{leuenhagen1993} and $T_\mathrm{eff} = 117\,000 \pm 5000$\,K and $\log g = 5.5$ \citep{toala2015,werner1992}.
Their element mass fractions are also included in \ta{tab:comp}}.
Obvious are the higher He and lower C, N, and O abundances in our two hybrid PG\,1159 stars. 
This may result from different evolutionary channels. The CSs of \object{Abell\,30} and \object{Abell\,78}
both underwent a born-again scenario \citep{ibenetal1983} resulting in a return to the
AGB, whereas the hybrid PG\,1159 stars experience a final He-shell flash at the departure
from the AGB.
This AFTP evolution may be the reason for the smaller amount of C, N, and O in the atmosphere in contrast to a (V)LTP scenario. 
\citet{toala2015} found a Fe deficiency of about one dex for \object{Abell\,78}.
This subsolar Fe abundance is in good agreement with the results for \wda and \wdn, 
although they show a smaller Fe deficiency. 
The high Ne abundance of $4$\,\% by mass  of the CS of \object{Abell\,78} \citep{toala2015}
and the revised N abundance of $1.5$\,\% by mass for both CSPNe \citep{guerrero2012,toala2015}
exceed the values determined for out two hybrid stars and are also an 
indicator for different evolutionary channels, namely VLTP and AFTP evolution. 
In common with the CS of \object{Abell\,78} is the high abundance of F 
\citep[$25$ times solar, ][]{toala2015} compared to $9.3$ and $29$ times solar, for \wda and \wdn respectively).
Their mass loss rates of 
$\dot{M} / M_{\odot} = 2.0 \times 10^{-8}\,\mathrm{yr}^{-1}$ 
\citep{guerrero2012} and $\dot{M} / M_{\odot} = 1.6 \times 10^{-8}\,\mathrm{yr}^{-1}$ 
\citep{toala2015} are about a factor two higher than the ones determined for \wda and \wdn \sK{sect:powr}.
The PNe \object{Abell\,30} and \object{Abell\,78} look very similar in shape but appear different to the 
``Galactic Soccerballs''.

\subsection{Comparison to PG\,1159 stars, hot post-AGB stars, and nucleosynthesis models}
\label{subsect:compy}
\citet{karakas2016} presented a grid of evolutionary models for different initial masses and 
metallicities. For stars with initial masses $M \le 3\,M_\odot$, they predict an enhanced 
production of C, N, F, Ne, and Na compared to solar values and normalized to the value for Fe. 
This prediction for the surface abundances of post-AGB stars is in line with our abundance determinations \sA{fig:x}. 
Another set of evolutionary models for initial masses of $1.8 - 6\,M_\odot$ was calculated 
by \citet{shingles2013} to investigate the resulting element yields of the species 
He, C, O, F, Ne, Si, P, S, and Fe depending on uncertainties in nucleosynthesis processes. They present the intershell abundances of their stellar models that should represent the surface abundances of PG\,1159 stars. As described in \sK{sect:intro}, the surface abundances of hybrid PG\,1159 stars should reflect a mixture of the abundances of the former H-rich envelope and the intershell.\\
The C abundances of our two program stars resemble the values of other PG\,1159 stars 
like for example the prototype star \object{PG\,1159$-$035} \citep{jahnetal2007}, the hot PG\,1159 stars
\object{PG\,1520+525} and \object{PG\,1144+005} \citep{werneretal2016} and the ``cooler'' 
\object{PG\,1707+427} and \object{PG\,1424+535} \citep{werner2015}. \\
The subsolar O abundances of the two stars analyzed in this work are significantly lower than 
the supersolar values of the stars mentioned above. 
However, the O abundance of the hybrid PG\,1159 stars lie within the predicted range of 
\citet{shingles2013} (O mass fraction between 0.2 and 1.2\,\%) and \citet{lawlormacdonald2006} in comparison to the PG\,1159 stars.
The large range in O abundances may be caused by different effectiveness of convective boundary mixing of the pulse-driven 
convection zone into the C/O core in the thermal pulses on the AGB.\\
An N enrichment is predicted for PG\,1159 stars that experienced a VLTP scenario due to 
a large production of $^{14}$N in an H-ingestion flash \citep[HIF, cf., ][]{wernerherwig2006}.
\wda and \wdn experienced an AFTP without HIF what corresponds to their comparatively low N
abundance. The value for \wdn lies within the range that is predicted by AFTP models whereas the value for \wda is higher.\\
The set of PG\,1159 stars shows a large range of F abundances from low mass fractions of $3.2 \times 10^{-6}$ for the prototype \object{PG\,1159$-$035} to values of $1.0 \times 10^{-4}$ for \object{PG\,1707$-$427}. The values of \wda and \wdn 
lie within this range but below the predictions of \citet{shingles2013} (F mass fractions between $2.0 \times 10^{-5}$ and $2.7 \times 10^{-4}$). The F production 
in the intershell region is very sensitive on the temperature and therefore on the 
initial mass \citep{lugaro2004}. They predict the highest F abundances for stars with 
initial masses of $2-4\,M_\odot$ at solar metallicity (F intershell mass fractions of $3-7 \times 10^{-5}$). Again, the values of \wdn and \wda are about 3 to 10 times lower than these predictions.\\
The Ne abundances for the set of mentioned PG\,1159 stars are all supersolar and in agreement with evolutionary models.
The determined values for the Si, P, and S abundances are at the lower border or 
slightly below the predictions from evolutionary models. The same holds for the 
solar upper abundance limit for P in PG\,1159$-$035.\\
The Fe deficiency is are not explained by nucleosynthesis calculations like those of \citet{karakas2016} that predict solar 
Fe abundances for stars with a initial solar composition, which is due to the fact that iron is not strongly depleted in nuclear processes in AGB stars with masses ranging from $0.8-8.0\,M_\odot$ as it is the case for the precursor AGB stars of \pna and \pnn. The models of \citet{shingles2013} yield a slightly subsolar Fe abundances but still cannot explain the observed deficiency. 
\textcolor{black}{For some of the PG\,1159 stars of the set used for comparison here, the Fe abundance has been measured and all are found to be solar. The only other star in \ab{fig:mfall} which, besides \wda and \wdn, shows a Fe deficiency is the CSPN of Abell\,78 \sK{subsect:A3078}. The Fe deficiency of 0.4 and 0.8\,dex for \wda and \wdn, respectively, does not include solar values within the error ranges.}
\citet{loebling2018} discuss the speculation of \citet{herwigetal2003} that this underabundance can be caused by neutron capture during the former AGB phase leading 
to a transformation of Fe into Ni and heavier elements. 
Probably the low Fe abundances determined in \wda and \wdn are just a consequence of a low initial metal content for these objects. \textcolor{black}{This subsolar metallicity does not rule out a thick or even thin disc membership \citep{reico2014}. Also their location and distance to the Galactic plane support that these are disc objects.} The models of \citet{karakas2016} predict a strong enhancement of the TIEs 
in the atmospheres of post-AGB stars.
Unfortunately, no abundance determination was possible for the TIEs
but due to the determined upper abundance limits for Zn, Ga, Ge, Kr, and Te 
of both stars and Zr for \wdn a strong enhancement can be ruled out. This is consistent with our determination of the wind intensity for which the wind is coupled. Thus, this prevents diffusion and disrupts the equilibrium balance between radiative levitation and gravitational settling.
To improve the analysis of TIEs, spectra with much better S/N and the calculation of
reliable atomic data are highly desirable.

\section{Conclusion}
\label{sect:conclusion}
{Regarding the evolutionary scenario for \wda and \wdn, an AFTP remains the best available candidate. The AFTP models presented here (Fig. \ref{fig:logglogteff} and Table \ref{tab:lawlor}) are able to reproduce qualitatively the observed trends. Abundances of H, He, C, N, and Ne are reasonably well reproduced by AFTP models. However AFTP models computed from sequences that include convective boundary mixing at the bottom of the pulse driven convective zone (as those presented here) display O abundances much larger than those of our program stars. In fact, the O abundances shown by our stars are in good agreement with those predicted by \cite{lawlormacdonald2006}, using AGB models that do not include any type of convective boundary mixing at the bottom of the pulse driven convective zone. These models, however, underestimate the C and Ne abundance, while the H and He abundances reproduce the observed values. Yet, the main argument against the AFTP scenario comes from the expansion ages of the nebulae. According to stellar evolution computations, AFTP models reach the location of our program stars in the HRD less than 2000\,yr after departing from the AGB, while the expansion ages are several times larger than this. This is true regardless the mass-loss prescription adopted in the post-AGB evolution. Unless the masses of the CSPNe are much lower than derived here, the AFTP scenario is unable to reproduce this key observational feature.\\
It is desirable to determine the iron abundance and the abundances of 
heavier elements of the ambient PN
to investigate on the photospheric composition of \pna and \pnn\ at the time of the PN's ejection
and to analyze the elements produced and ejected during the preceding AGB phase \citep[cf., ][]{lugaro2017} 
but this is out of the focus of this paper. 

\section*{Acknowledgements}
LL is supported by the German Research Foundation (DFG, grant WE\,1312/49-1).
M3B is supported by the PICT 2016-0053 from ANPCyT, Argentina.
This work was partially funded by DA/16/07 grant form DAAD-MinCyT bilateral cooperation program.
We were supported by the High Performance and Cloud Computing Group 
at the Zentrum f\"ur Datenverarbeitung of the University of T\"ubingen, the state of 
Baden-W\"urttemberg through bwHPC, and the DFG (grant INST\,37/935-1\,FUGG).
The GAVO project had been supported by the Federal Ministry of Education and Research (BMBF) 
at T\"ubingen (05\,AC\,6\,VTB, 05\,AC\,11\,VTB).
We thank the referee Ralf Napiwotzki for his many useful comments that improved this paper.
This work used the profile-fitting procedure OWENS developed by M\@. Lemoine and the FUSE French Team.
We thank Falk Herwig, Timothy Lawlor, and James MacDonald for helpful comments and discussion,
Ralf Napiwotzki for putting the reduced VLT spectra at our disposal.
The 
TIRO (\url{http://astro-uni-tuebingen.de/~TIRO}),
TMAD (\url{http://astro-uni-tuebingen.de/~TMAD}), and
TOSS (\url{http://astro-uni-tuebingen.de/~TOSS}) services used for this paper 
were constructed as part of the activities of the German Astrophysical Virtual Observatory.
Some of the data presented in this paper were obtained from the
Mikulski Archive for Space Telescopes (MAST). STScI is operated by the
Association of Universities for Research in Astronomy, Inc., under NASA
contract NAS5-26555. Support for MAST for non-HST data is provided by
the NASA Office of Space Science via grant NNX09AF08G and by other
grants and contracts. 
This research has made use of 
NASA's Astrophysics Data System and
the SIMBAD database, operated at CDS, Strasbourg, France.
This work has made use of data from the European Space Agency (ESA) mission
{\it Gaia} (\url{https://www.cosmos.esa.int/gaia}), processed by the {\it Gaia}
Data Processing and Analysis Consortium (DPAC,
\url{https://www.cosmos.esa.int/web/gaia/dpac/consortium}). Funding for the DPAC
has been provided by national institutions, in particular the institutions
participating in the {\it Gaia} Multilateral Agreement.




\bibliographystyle{mnras}
\bibliography{a43ngc7094}

\begin{figure*}
\color{black}
  \resizebox{\hsize}{!}{\includegraphics{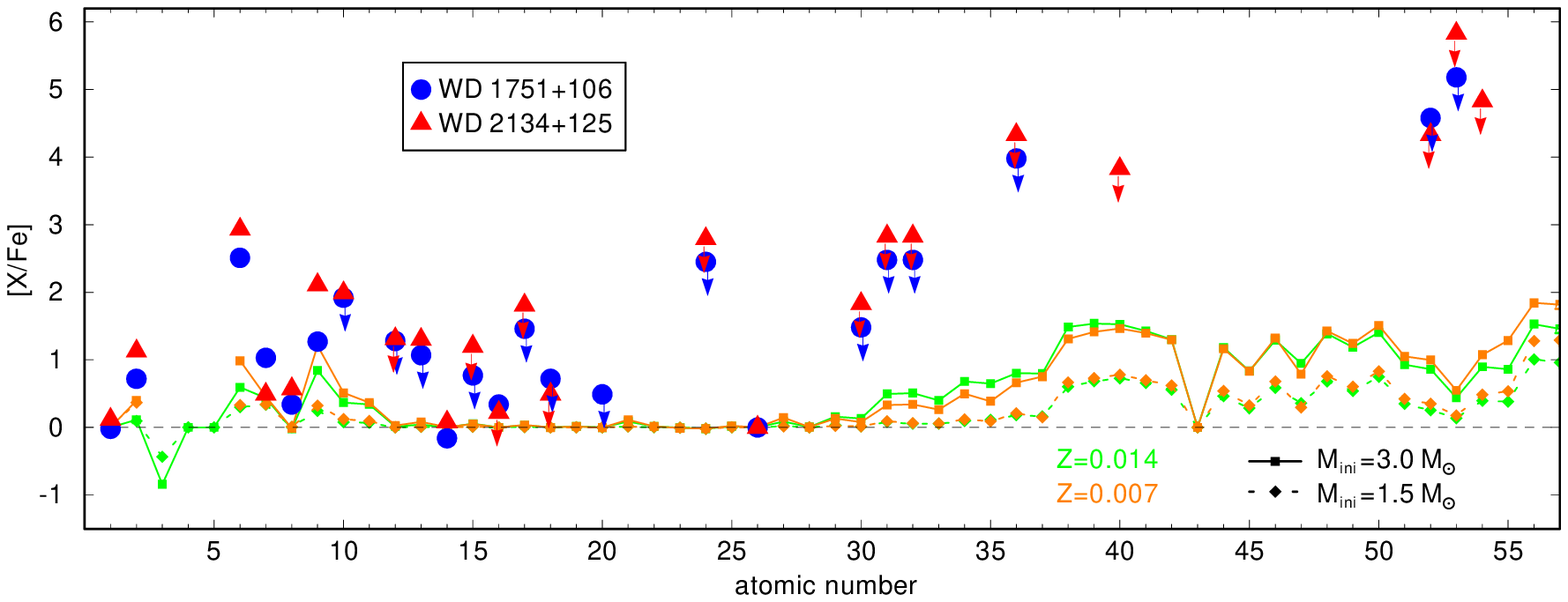}}
   \caption{Photospheric abundance ratio $[\mathrm{X}/\mathrm{Fe}] = \log (n_\mathrm{X} / n_\mathrm{Fe} ) - \log (n_{\mathrm{X},\odot} / n_{\mathrm{Fe},\odot} )$ with the number fraction $n_\mathrm{X}$ for element X of \wda and \wdn  determined from detailed line-profile fits. Upper limits are indicated with arrows.
            Predictions of \citet[][$M_\mathrm{ini} = 1.5$ and $3.0$ (see legend); metallicity $Z=0.014$ (green) and $0.007$ (cyan)]{karakas2016} are shown for comparison.}
   \label{fig:x}
\end{figure*}
  
\begin{table*}
\setlength{\tabcolsep}{0.38em}
\caption[]{Parameters of \wda and \wdn compared with literature values.}
\centering
\label{tab:finab}
\begin{tabular}{rrrcrrcrrcrr}
\hline
\hline
\noalign{\smallskip}
 & \multicolumn{5}{c}{\wda} && \multicolumn{5}{c}{\wdn} \\
\cline{2-6} \cline{8-12}
\noalign{\smallskip}
& \multicolumn{2}{c}{Literature} && \multicolumn{2}{c}{This work} && \multicolumn{2}{c}{Literature} && \multicolumn{2}{c}{This work}\\
\cline{2-3} \cline{5-6} \cline{8-9} \cline{11-12}
\noalign{\smallskip}
$T_\mathrm{eff}$\,/\,kK      & \multicolumn{2}{c}{$115\pm 5$\,$^{a}$} && 
                              \multicolumn{2}{c}{$115\pm 5$}                   && 
                              \multicolumn{2}{c}{$115\pm 5$\,$^{a}$} && 
                              \multicolumn{2}{c}{$115\pm 5$}                   \\
$\log\,(g$\,/\,cm/s$^2$)    & \multicolumn{2}{c}{\hspace{1.5mm}$5.5\pm 0.1$\,$^{a}$} && 
                              \multicolumn{2}{c}{\hspace{3.5mm}$5.6\pm 0.1$}                  && 
                              \multicolumn{2}{c}{\hspace{2.0mm}$5.4\pm 0.1$\,$^{a}$} && 
                              \multicolumn{2}{c}{\hspace{3.5mm}$5.6\pm 0.1$}                  \\
$E_\mathrm{B-V}$             & \multicolumn{2}{c}{$0.265\pm 0.035$\,$^{e}$} && 
                              \multicolumn{2}{c}{$0.265\pm 0.010$} && 
                              \multicolumn{2}{c}{$0.150\pm 0.040$\,$^{d}$} && 
                              \multicolumn{2}{c}{$0.135\pm 0.010$} \\
$N_\ion{H}{I}$\,/\,cm$^{-2}$ & \multicolumn{2}{c}{$(1.0\pm 0.2)\times 10^{21}$\,$^{e}$} && 
                             \multicolumn{2}{c}{$(1.0\pm 0.1)\times 10^{21}$} && 
                             \multicolumn{2}{c}{$(7.0\pm 0.1)\times 10^{20}$\,$^{d}$} && 
                             \multicolumn{2}{c}{$(6.5\pm 0.1)\times 10^{20}$} \\
$v_\mathrm{rad}$\,/\,km/s    & \multicolumn{2}{c}{$-42.0\pm 11.5$\,$^{c}$} && 
                              \multicolumn{2}{c}{$-100\pm 10$} && 
                              \multicolumn{2}{c}{$-101.1\pm 30.8$\,$^{c}$} && 
                              \multicolumn{2}{c}{$-53\pm 10$} \\
$d$\,/\,kpc                  & \multicolumn{2}{c}{$2.47\pm 0.30$\,$^{f}$} && 
                              \multicolumn{2}{c}{$2.23^{+0.31}_{-0.33}$} && 
                              \multicolumn{2}{c}{$1.75\pm 0.36$\,$^{f}$} && 
                              \multicolumn{2}{c}{$1.65^{+0.32}_{-0.31}$} \\
\noalign{\smallskip}
$M\,/\,M_\odot$             &  \multicolumn{2}{c}{$0.53^{+0.10}_{-0.02}$\,$^{e}$} && 
                              \multicolumn{2}{c}{$0.57^{+0.07}_{-0.04}$} && 
                              \multicolumn{2}{c}{$0.53^{+0.06}_{-0.06}$\,$^{d}$} && 
                              \multicolumn{2}{c}{$0.57^{+0.07}_{-0.04}$} \\
\noalign{\smallskip}
$\log\ ( L\,/\,L_\odot )$   &  \multicolumn{2}{c}{$3.44^{+0.50}_{-0.58}$\,$^{e}$} && 
                              \multicolumn{2}{c}{$3.77^{+0.23}_{-0.24}$} && 
                              \multicolumn{2}{c}{$ $} && 
                              \multicolumn{2}{c}{$3.77^{+0.23}_{-0.24}$} \\ 
\noalign{\smallskip}
$R_\mathrm{PN}$\,/\,pc      &  \multicolumn{2}{c}{$0.51$\,$^{h}$} && 
                              \multicolumn{2}{c}{$0.48^{+0.07}_{-0.07}$} && 
                              \multicolumn{2}{c}{$0.51$\,$^{h}$} && 
                              \multicolumn{2}{c}{$0.42^{+0.08}_{-0.08}$} \vspace{0.5mm}\\
\hline
\end{tabular}
\newline
\textbf{Notes. }
$^{(a)}${\citet{loebling2018}},
$^{(b)}${\citet{ziegleretal2009a}},
$^{(c)}${\citet{durandetal1998}},
$^{(d)}${\citet{ziegler2008}},
$^{(e)}${\citet{friederich2010}},
$^{(f)}${\citet{frewetal2016}},
$^{(g)}${\citet{ringatetal2011}},
$^{(h)}${\citet{napiwotzki1999}}

\end{table*}
\addtocounter{table}{-1}
\begin{landscape}
\centering
\begin{table}
\color{black}
\centering
\setlength{\tabcolsep}{0.4em}
\caption[]{Continued.}
\begin{tabular}{rrrcrrrrcrrcrrrr}
\hline
\hline
\noalign{\smallskip}
 & \multicolumn{7}{c}{\wda} && \multicolumn{7}{c}{\wdn} \\
\cline{2-8} \cline{10-16}
\noalign{\smallskip}
& \multicolumn{2}{c}{Literature} && \multicolumn{4}{c}{This work} && \multicolumn{2}{c}{Literature} && \multicolumn{4}{c}{This work}\\
\cline{2-3} \cline{5-8} \cline{10-11} \cline{13-16}
\noalign{\smallskip}
\noalign{\smallskip}
Abundances & [X] & Mass fraction && [X] & Mass fraction & Number fraction & [X/Fe] && [X] & Mass fraction && [X] & Mass fraction & Number fraction & [X/Fe]\\
\cline{2-3} \cline{5-8} \cline{10-11} \cline{13-16}
\noalign{\smallskip}
H  &   $-$ 0.483\,$^{g}$ &       $ 2.43\times 10^{-1}$ && $-$   0.5    &   $2.5  \times 10^{-1}$   & $   6.4  \times 10^{-1}$  &  $-  $0.0       &&   $-$ 0.62\,$^{b}$    &  $ 1.77\times 10^{-1}$      && $-$    0.7  & $1.5  \times 10^{-1}$     &    $   4.9  \times 10^{-1}$  &                 0.1                 \\
He &       0.350\,$^{g}$ &       $ 5.59\times 10^{-1}$ &&       0.3    &   $4.6  \times 10^{-1}$   & $   3.0  \times 10^{-1}$  &  $   $0.7       &&     0.45\,$^{b}$      &  $ 7.03\times 10^{-1}$      &&        0.3  & $5.2  \times 10^{-1}$     &    $   4.3  \times 10^{-1}$  &                 1.1                    \\
C  &       1.909\,$^{g}$ &       $ 1.75\times 10^{-1}$ &&       2.1    &   $2.7  \times 10^{-1}$   & $   5.8  \times 10^{-2}$  &  $   $2.5       &&     1.76\,$^{b}$      &  $ 1.24\times 10^{-1}$      &&        2.1  & $3.1  \times 10^{-1}$     &    $   8.4  \times 10^{-2}$  &                 2.9                    \\
N  &   $-$ 0.491\,$^{g}$ &       $ 2.00\times 10^{-4}$ &&       0.6    &   $2.6  \times 10^{-3}$   & $   4.8  \times 10^{-4}$  &  $   $1.0       &&   $-$ 0.85\,$^{b}$    &  $ 8.73\times 10^{-5}$      && $-$    0.3  & $3.3  \times 10^{-4}$     &    $   7.7  \times 10^{-5}$  &                 0.5                \\
O  &   $-$ 0.516\,$^{g}$ &       $ 1.63\times 10^{-3}$ && $-$   0.1    &   $4.4  \times 10^{-3}$   & $   7.2  \times 10^{-4}$  &  $   $0.3       &&   $-$ 1.81\,$^{b}$    &  $ 8.20\times 10^{-5}$      && $-$    0.2  & $3.2  \times 10^{-3}$     &    $   6.7  \times 10^{-4}$  &                 0.6                \\
F  &       0.694\,$^{g}$ &       $ 2.50\times 10^{-6}$ &&       1.0    &   $3.3  \times 10^{-6}$   & $   4.5  \times 10^{-7}$  &  $   $1.3       &&     0.34\,$^{b}$      &  $ 1.11\times 10^{-6}$      &&        1.5  & $9.9  \times 10^{-6}$     &    $   1.7  \times 10^{-6}$  &                 2.1                    \\
Ne &                     &                            && $\le$ 1.5    &$\le3.6  \times 10^{-2}$   & $\le4.7  \times 10^{-3}$  &  $\le$ 1.9      &&     0.00\,$^{b}$      &  $ 1.02\times 10^{-3}$      &&        1.2  & $1.9  \times 10^{-3}$     &    $   3.1  \times 10^{-3}$  &                  2.0                    \\
Mg & $\le$ 0.108\,$^{e}$ & $\le$ $ 7.74\times 10^{-4}$ && $\le$ 0.8    &$\le4.7  \times 10^{-3}$   & $\le5.0  \times 10^{-4}$  &  $\le$1.3       &&                       &                            && $\le$  0.5  & $\le2.2  \times 10^{-3}$  & $\le  3.0  \times 10^{-4}$  &          $\le$    1.3                    \\
Al & $\le$ 0.797\,$^{e}$ & $\le$ $ 2.90\times 10^{-4}$ && $\le$ 0.6    &$\le2.3  \times 10^{-4}$   & $\le2.2  \times 10^{-5}$  &  $\le$1.1       &&                       &                            &&        0.5  & $1.7  \times 10^{-4}$     &    $   2.1  \times 10^{-5}$  &                  1.3                    \\
Si &   $-$ 0.560\,$^{g}$ &       $ 1.83\times 10^{-4}$ && $-$   0.6    &   $1.6  \times 10^{-4}$   & $   1.5  \times 10^{-5}$  &  $-  $0.2       &&   $-$ 0.21\,$^{b}$    &  $ 4.10\times 10^{-4}$      && $-$    0.7  & $1.2  \times 10^{-4}$   &    $   1.4  \times 10^{-5}$  &                   0.1               \\
P  &   $-$ 0.593\,$^{g}$ &       $ 1.33\times 10^{-6}$ &&$\le$  0.3    &$\le1.2  \times 10^{-5}$   & $\le1.0  \times 10^{-6}$  &  $\le$0.8       &&   $-$ 1.15\,$^{b}$    &  $ 3.70\times 10^{-7}$      && $\le$  0.4  & $\le1.4  \times 10^{-5}$  & $\le  1.5  \times 10^{-6}$  &          $\le$   1.2                    \\
S  &   $-$ 0.378\,$^{g}$ &       $ 1.36\times 10^{-4}$ && $-$   0.1    &   $2.4  \times 10^{-4}$   & $   1.9  \times 10^{-5}$  &  $   $0.3       &&     0.16\,$^{b}$      &  $ 4.69\times 10^{-4}$      && $\le -$ 0.6 & $\le7.8  \times 10^{-5}$  & $\le  8.0  \times 10^{-6}$  &          $\le$   0.2                     \\
Cl &                    &                             && $\le$ 1.0    &$\le8.3  \times 10^{-5}$   & $\le6.1  \times 10^{-6}$  &  $\le$ 1.5      &&                       &                            && $\le$  1.0  & $\le8.1  \times 10^{-5}$  & $\le  7.5  \times 10^{-6}$  &          $\le$     1.8                    \\
Ar & $\le$ 1.307\,$^{e}$ & $\le$ $ 1.04\times 10^{-3}$ && $\le$ 0.3    &$\le1.4  \times 10^{-4}$   & $\le8.8  \times 10^{-6}$  &  $\le$0.7       &&     0.00\,$^{d}$      &  $ 4.43\times 10^{-5}$      && $\le -$ 0.3  & $\le3.5  \times 10^{-5}$  & $\le  2.9  \times 10^{-6}$  &         $\le$   0.5                      \\
Ca &                     &                            && $\le$ 0.0    &$\le6.5  \times 10^{-5}$   & $\le4.5  \times 10^{-6}$  &  $\le$ 0.5      &&                       &                            &&             &                           &                              &                                      \\
Cr &                     &                            && $\le$ 2.0    &$\le1.7  \times 10^{-3}$   & $\le8.1  \times 10^{-5}$  &  $\le$ 2.5      &&                       &                            && $\le$  2.0  & $\le1.6  \times 10^{-3}$  & $\le  1.0  \times 10^{-4}$  &          $\le$   2.8                      \\
Fe &   $-$ 0.691\,$^{e}$ &       $ 2.35\times 10^{-4}$ && $-$   0.4    &   $4.5  \times 10^{-4}$   & $   2.1  \times 10^{-5}$  &  $   $0.0       && $\le -$ 1.00\,$^{b}$  & $\le$ $ 1.15\times 10^{-4}$ && $-$    0.8  & $2.0  \times 10^{-4}$     &  $      1.2  \times 10^{-5}$ &           0.0               \\
Ni &$\le -$ 1.000\,$^{e}$& $\le$ $ 7.30\times 10^{-6}$ &&              &                          &                           &  $   $       && $\le -$ 1.00\,$^{b}$  & $\le$ $ 7.30\times 10^{-6} $   &&            &                           &                              &                                         \\
Zn &                    &                             && $\le$ 1.0   &$\le1.7  \times 10^{-5}$   & $\le7.2  \times 10^{-7}$  &  $\le$1.5       &&                       &                            && $\le$  1.0  & $\le1.7  \times 10^{-5}$  & $\le  9.1  \times 10^{-7}$  &          $\le$   1.8                      \\
Ga &                    &                             && $\le$ 2.0   &$\le5.6  \times 10^{-6}$   & $\le2.2  \times 10^{-7}$  &  $\le$2.5       &&                       &                            && $\le$  2.0  & $\le5.6  \times 10^{-6}$  & $\le  2.8  \times 10^{-7}$  &          $\le$   2.8                      \\
Ge &                    &                             && $\le$ 2.0   &$\le2.4  \times 10^{-5}$   & $\le8.9  \times 10^{-7}$  &  $\le$2.5       &&                       &                            && $\le$  2.0  & $\le2.4  \times 10^{-5}$  & $\le  1.1  \times 10^{-6}$  &          $\le$   2.8                      \\
Kr &                    &                             && $\le$ 3.5   &$\le3.4  \times 10^{-4}$   & $\le1.1  \times 10^{-5}$  &  $\le$4.0       &&                       &                            && $\le$  3.5  &$\le3.4  \times 10^{-4}$   & $\le  1.4  \times 10^{-5}$  &          $\le$   4.3                      \\
Zr &                    &                             &&             &                           &                          &  $   $       &&                       &                               && $\le$  3.0 &$\le2.5  \times 10^{-5}$   &  $\le 9.6  \times 10^{-7}$  &            $\le$    3.8                      \\
Te &                    &                             && $\le$ 4.0   &$\le1.4  \times 10^{-4}$   & $\le3.8  \times 10^{-6}$  &  $\le$4.6       &&                       &                            && $\le$  3.5 &$\le4.5  \times 10^{-5}$   &  $\le 1.2  \times 10^{-6}$  &           $\le$   4.3                      \\
I  &                    &                             && $\le$ 4.6   &$\le1.3  \times 10^{-4}$   & $\le3.6  \times 10^{-6}$  &  $\le$5.2       &&                       &                            && $\le$  5.0 &$\le3.3  \times 10^{-4}$   &  $\le 8.9  \times 10^{-6}$  &           $\le$   5.8                      \\
Xe &                    &                             &&              &                           &      		     &              &&                       &                               && $\le$  4.0 &$\le1.7  \times 10^{-4}$   &  $\le 4.4  \times 10^{-6}$  &            $\le$    4.8                      \\
\hline
\end{tabular}
\newline
\textbf{Notes. }
[X] = log\,(abundance/solar abundance), $[\mathrm{X}/\mathrm{Fe}] = \log (n_\mathrm{X} / n_\mathrm{Fe} ) - \log (n_{\mathrm{X},\odot} / n_{\mathrm{Fe},\odot} )$ with the number fraction $n_\mathrm{X}$ for element X, 
the error of our abundance determination is $\pm 0.3$\,dex,
$^{(a)}${\citet{loebling2018}},
$^{(b)}${\citet{ziegleretal2009a}},
$^{(c)}${\citet{durandetal1998}},
$^{(d)}${\citet{ziegler2008}},
$^{(e)}${\citet{friederich2010}},
$^{(f)}${\citet{frewetal2016}},
$^{(g)}${\citet{ringatetal2011}},
$^{(h)}${\citet{napiwotzki1999}}
\end{table}
\end{landscape}


\appendix

\section{Additional figures and tables.}
\label{app:additional}
\FloatBarrier

\begin{table*}
\centering
\caption{Statistics of the H -- Ar$^{a}$ and Ca - Ba$^{b}$ model atoms used in our model-atmosphere calculations.}
\label{tab:stat}
\setlength{\tabcolsep}{.3em}
\begin{tabular}{rlrrrp{10mm}rlrrr}
\hline
\hline
\multicolumn{2}{l}{}    & \multicolumn{2}{c}{Levels} & & &\multicolumn{2}{l}{}& \multicolumn{1}{l}{Super}                   & \multicolumn{1}{c}{Super} & Individual\\
\cline{3-4}
\multicolumn{2}{l}{}    &      &        &          && \multicolumn{2}{l}{}    &                         &        &       \vspace{-5.5mm}\\
\multicolumn{2}{l}{\hbox{}\hspace{4mm}Ion} &      &        &  ~Lines  && 
\multicolumn{2}{l}{\hbox{}\hspace{4mm}Ion} &        &        &       \vspace{-1.5mm}\\
\multicolumn{2}{l}{}    & NLTE & ~~~LTE &          && \multicolumn{2}{l}{}    & \multicolumn{1}{c}{levels$^{c}$} & \multicolumn{1}{c}{lines}  & \multicolumn{1}{c}{lines}                \\
\hline
\noalign{\smallskip}
 H & \Ion{}{1} &   10 &   22 &    45 && Ca                  & \Ion{}{7} &    7 &    27 &      71\,608 \\
   & \Ion{}{2} &    1 &    0 &   $-$ &&                     & \Ion{}{8} &    7 &    26 &       9\,124 \\
He & \Ion{}{1} &    5 &   98 &     3 &&                     & \Ion{}{9} &    1 &     0 &            0 \\
   & \Ion{}{2} &   16 &   16 &   120 && Cr                  & \Ion{}{7} &    7 &    24 &      37\,070 \\
   & \Ion{}{3} &    1 &    0 &   $-$ &&                     & \Ion{}{8} &    7 &    25 &     132\,221 \\
 C & \Ion{}{3} &    1 &  104 &     0 &&                     & \Ion{}{9} &    1 &     0 &            0 \\
   & \Ion{}{4} &   54 &    4 &   295 && Fe                  & \Ion{}{7} &    7 &    24 &     200\,455 \\
   & \Ion{}{5} &    1 &    0 &     0 &&                     & \Ion{}{8} &    7 &    27 &      19\,587 \\
 N & \Ion{}{4} &    1 &   93 &     0 &&                     & \Ion{}{9} &    1 &     0 &            0 \\
   & \Ion{}{5} &   54 &    8 &   297 && IG$^{d}$ & \Ion{}{7} &    7 &    27 &  5\,216\,215 \\
   & \Ion{}{6} &    1 &    0 &     0 &&                     & \Ion{}{8} &    7 &    28 &  2\,218\,561 \\
 O & \Ion{}{5} &   10 &  124 &    14 &&                     & \Ion{}{9} &    1 &     0 &            0 \\
   & \Ion{}{6} &   54 &    8 &   291 && Zn                  & \Ion{}{5} &    7 &    15 &       1\,879 \\
   & \Ion{}{7} &    1 &    0 &     0 &&                     & \Ion{}{6} &    1 &     0 &            0 \\
 F & \Ion{}{5} &   27 &  101 &    81 && Ga                  & \Ion{}{5} &    7 &    15 &          517 \\
   & \Ion{}{6} &   35 &  105 &   119 &&                     & \Ion{}{6} &    7 &    13 &       1\,914 \\
   & \Ion{}{7} &    1 &    0 &     0 &&                     & \Ion{}{7} &    1 &     0 &            0 \\
Ne & \Ion{}{5} &   19 &   75 &    33 && Ge                  & \Ion{}{5} &    7 &    16 &       2\,159 \\
   & \Ion{}{6} &   31 &    0 &    73 &&                     & \Ion{}{6} &    7 &    12 &          414 \\
   & \Ion{}{7} &   36 &   73 &   132 &&                     & \Ion{}{7} &    1 &     0 &            0 \\
   & \Ion{}{8} &    1 &    0 &     0 && Se                  & \Ion{}{5} &    7 &    19 &          310 \\
Na & \Ion{}{5} &   23 &  301 &    49 &&                     & \Ion{}{6} &    1 &     0 &            0 \\
   & \Ion{}{6} &   28 &  365 &    69 &&                     & \Ion{}{7} &    1 &     0 &            0 \\
   & \Ion{}{7} &    1 &    0 &     0 && Kr                  & \Ion{}{6} &    7 &    19 &          843 \\
Mg & \Ion{}{5} &   21 &   31 &    31 &&                     & \Ion{}{7} &    7 &    21 &          743 \\
   & \Ion{}{6} &   27 &    0 &    60 &&                     & \Ion{}{8} &    1 &     0 &            0 \\
   & \Ion{}{7} &    1 &    0 &     0 && Sr                  & \Ion{}{6} &    7 &    10 &           70 \\
Al & \Ion{}{5} &   22 &  207 &    48 &&                     & \Ion{}{7} &    7 &    10 &           46 \\
   & \Ion{}{6} &   20 &  300 &    30 &&                     & \Ion{}{8} &    1 &     0 &            0 \\
   & \Ion{}{7} &    1 &    0 &     0 && Zr                  & \Ion{}{6} &    7 &    12 &       1\,098 \\
Si & \Ion{}{4} &   12 &   26 &    24 &&                     & \Ion{}{7} &    7 &    15 &          947 \\
   & \Ion{}{5} &   15 &   10 &    20 &&                     & \Ion{}{8} &    1 &     0 &            0 \\
   & \Ion{}{6} &    1 &    0 &     0 && Mo                  & \Ion{}{6} &    7 &    23 &          984 \\
 P & \Ion{}{5} &   25 &   12 &    49 &&                     & \Ion{}{7} &    7 &    16 &       1\,173 \\
   & \Ion{}{6} &   15 &    0 &     5 &&                     & \Ion{}{8} &    1 &     0 &            0 \\
   & \Ion{}{7} &    1 &    0 &     0 && Te                  & \Ion{}{5} &    1 &     0 &            0 \\
 S & \Ion{}{5} &    1 &  109 &     0 &&                     & \Ion{}{6} &    7 &    12 &          178 \\
   & \Ion{}{6} &   25 &   12 &    49 &&                     & \Ion{}{7} &    1 &     0 &            0 \\
   & \Ion{}{7} &    1 &    0 &     0 &&                     & \Ion{}{5} &    1 &     0 &            0 \\
Cl & \Ion{}{7} &   21 &    4 &    58 && I                   & \Ion{}{6} &    7 &    15 &          197 \\
   & \Ion{}{8} &    1 &    0 &     0 &&                     & \Ion{}{7} &    1 &     0 &            0 \\
Ar & \Ion{}{7} &   40 &  111 &   130 &&                     & \Ion{}{6} &    7 &    16 &          243 \\
   & \Ion{}{8} &   23 &   28 &    75 && Xe                  & \Ion{}{7} &    7 &    19 &          491 \\
   & \Ion{}{9} &    1 &    0 &     0 &&                     & \Ion{}{8} &    1 &     0 &            0 \\
   &           &      &      &       &&                     & \Ion{}{6} &    7 &     6 &          162 \\
   &           &      &      &       && Ba                  & \Ion{}{7} &    7 &    11 &          493 \\
   &           &      &      &       &&                     & \Ion{}{8} &    1 &     0 &            0 \\
\cline{1-11}
\noalign{\smallskip}
total &        &  694 & 2228 &  2199 &&                     &           &  206 &   478 &  7\,919\,702 \\
\hline
\end{tabular}

\textbf{Notes.} $^{(a)}${classical model atoms},
$^{(b)}${model atoms constructed using a statistical approach \citep{rauchdeetjen2003}},\\
$^{(c)}${treated as NLTE levels},
$^{(d)}${IG is a generic model atom \citep{rauchdeetjen2003} that includes opacities of Sc, Ti, V, Mn, Ni, and Co.}
\end{table*}

\begin{table*}
\begin{center}
\color{black}
\centering
\caption{Observation log for \wda and \wdn.}
\label{tab:obslog}
\vspace*{-2mm}
\begin{tabular}{lllcrrrr}
\hline
\hline
 &  &  &  & Wavelength  & Aperture/ & Exp.  & Resolving power\\
\vspace{-0.52cm}\\
Object & Instrument & Dataset Id & Start Time (UT) &  &  &  & \\
\vspace{-0.52cm}\\
 &  &  & & range ($\lambda$) & Grating & time (s) &$R = \lambda\,/\,\Delta\lambda$\\
\hline
\wda & FUSE\,$^\mathrm{a}$ & B0520201000 & 2001-07-29 20:41:47 & $917 - 1181$ & LWRS & $11\,438$ & 20\,000\\
  & FUSE & B0520202000 & 2001-08-03 22:18:20 & $917 - 1181$ & LWRS & $9\,528$ & 20\,000\\
  & GHRS\,$^\mathrm{b}$  & Z3GW0304T   & 1996-09-08 07:00:34 & $1\,140-1\,435$ & 2.0/ G140L & $4\,243$ & 2\,000\\
  & IUE\,$^\mathrm{c}$  & LWR08735    & 1980-09-06 21:45:21 & $1\,850-3\,300$ & LARGE & $3\,600$ & 10\,000\\
  & IUE  & SWP10245    & 1980-09-28 21:50:02 & $1\,150-2\,000$ & LARGE & $5\,100$ & 300\\
  & TWIN\,$^\mathrm{d}$ & & 2014-08-16 & $3\,500-5\,500$ & T08 & 1\,800 & 1\,500 \\
  & TWIN & & 2014-08-16 & $5\,500-7\,500$ & T04 & 1\,800 & 1\,500 \\
  & TWIN & & 2014-08-17 & $3\,500-5\,500$ & T08 & 1\,800 & 1\,500 \\
  & TWIN & & 2014-08-17 & $5\,500-7\,500$ & T04 & 1\,800 & 1\,500 \\
  & UVES\,$^\mathrm{e}$ & 167.D$-$0407(A) & 2001-06-18 05:02:50 & $3\,280-4\,560$ & Blue, CD2 & 300 & 18\,500\\
  & UVES & 167.D$-$0407(A) & 2001-07-26 01:26:34 & $3\,280-4\,560$ & Blue, CD2 & 300 & 18\,500\\
  & UVES & 167.D$-$0407(A) & 2001-06-18 05:03:38 & $4\,580-6\,690$ & Red, CD3 & 300 & 18\,500\\
  & UVES & 167.D$-$0407(A) & 2001-07-26 01:27:48 & $4\,580-6\,690$ & Red, CD3 & 300 & 18\,500\\
WD\,2134+125 & FUSE & P1043701000 & 2000-11-13 08:53:28 & $911-1\,188$ & LWRS & $22\,754$ & 20\,000\\
  & STIS\,$^\mathrm{f}$  & O8MU02010   & 2004-06-24 20:43:30 & $1\,150-1\,730$ & E140M & $650$ & 45\,800\\
  & STIS & O8MU02020   & 2004-06-24 22:19:29 & $1\,150-1\,730$ & E140M & $656$& 45\,800\\
  & STIS & O8MU02030   & 2004-06-24 23:55:29 & $1\,150-1\,730$ & E140M & $655$& 45\,800\\
  & STIS & OCY508010   & 2016-07-25 18:40:44 & $1\,680-3\,060$ &G230LB &200&700\\
  & STIS & OCY508020   & 2016-07-25 18:44:27 & $1\,680-3\,060$ &G230LB &200&700\\
  & STIS & OCY508030   & 2016-07-25 18:48:38 & $1\,680-3\,060$ &G230LB &200&700\\
  & STIS & OCY508040   & 2016-07-25 18:52:21 & $1\,680-3\,060$ &G230LB &200&700\\
  & STIS & OCY508F8Q   & 2016-07-25 18:58:39 & $2\,900-5\,700$ &G430L &70&500\\
  & STIS & OCY508F9Q   & 2016-07-25 19:00:10 & $2\,900-5\,700$ &G430L &70&500\\
  & STIS & OCY508FAQ   & 2016-07-25 19:02:09 & $2\,900-5\,700$ &G430L &70&500\\
  & STIS & OCY508FBQ   & 2016-07-25 19:03:40 & $2\,900-5\,700$ &G430L &70&500\\
  & STIS & OCY508FCQ   & 2016-07-25 19:08:02 & $5\,240-1\,0270$ &G750L &110&500\\
  & STIS & OCY508FDQ   & 2016-07-25 19:10:12 & $5\,240-1\,0270$ &G750L &110&500\\
  & STIS & OCY508FEQ   & 2016-07-25 19:12:50 & $5\,240-1\,0270$ &G750L &110&500\\
  & STIS & OCY508FFQ   & 2016-07-25 19:15:00 & $5\,240-1\,0270$ &G750L &110&500\\
  & IUE  & LWR10774    & 1981-06-03 22:51:10 & $1\,850-3\,300$ & LARGE & $1\,080$& 300\\
  & IUE  & LWR10909    & 1981-06-20 14:57:17 & $1\,850-3\,300$ & LARGE & $1\,800$& 300\\
  & IUE  & LWP23315    & 1992-06-17 15:57:41 & $1\,850-3\,300$ & LARGE & $1\,380$& 300\\
  & IUE  & SWP14289    & 1981-06-20 14:40:41 & $1\,150-2\,000$ & LARGE & $720$& 10\,000 \\
  & IUE  & SWP28248    & 1986-05-01 16:42:03 & $1\,150-2\,000$ & LARGE & $1\,440$& 10\,000\\
  & IUE  & SWP44943    & 1992-06-17 16:32:32 & $1\,150-2\,000$ & LARGE  & $900$& 10\,000\\
  & TWIN & & 2014-08-15 & $3\,500-5\,500$ & T08 & 1\,800 & 1\,500 \\
  & TWIN & & 2014-08-15 & $5\,500-7\,500$ & T04 & 1\,800 & 1\,500 \\
  & TWIN & & 2014-08-17 & $3\,500-5\,500$ & T08 & 1\,800 & 1\,500 \\
  & TWIN & & 2014-08-17 & $5\,500-7\,500$ & T04 & 1\,800 & 1\,500 \\
  & TWIN & & 2014-08-18 & $3\,500-5\,500$ & T08 & 1\,800 & 1\,500 \\
  & TWIN & & 2014-08-18 & $5\,500-7\,500$ & T04 & 1\,800 & 1\,500 \\
  & UVES & 167.D$-$0407(A) & 2001-08-21 01:58:34 & $3\,280-4\,560$ & Blue, CD2 & 300 & 18\,500\\
  & UVES & 167.D$-$0407(A) & 2001-09-01 00:49:29 & $3\,280-4\,560$ & Blue, CD2 & 300 & 18\,500\\
  & UVES & 167.D$-$0407(A) & 2001-09-20 01:38:14 & $3\,280-4\,560$ & Blue, CD2 & 300 & 18\,500\\
  & UVES & 167.D$-$0407(A) & 2001-08-21 02:00:03 & $4\,580-6\,690$ & Red, CD3& 300 & 18\,500\\
  & UVES & 167.D$-$0407(A) & 2001-09-01 00:51:03 & $4\,580-6\,690$ & Red, CD3& 300 & 18\,500\\
  & UVES & 167.D$-$0407(A) & 2001-09-20 01:33:09 & $4\,580-6\,690$ & Red, CD3& 300 & 18\,500\\
\hline
\end{tabular}
\end{center}
\vspace{-3mm}
\begin{footnotesize}
\color{black}
\raggedright{
\textbf{Notes.} 
a: \hspace{0.5mm} Far Ultraviolet Spectroscopic Explorer,\hspace{1.5mm}
b: \hspace{0.5mm} Goddard High-Resolution Spectrograph,\hspace{1.5mm}
c: \hspace{0.5mm} International Ultraviolet Explorer,\hspace{1.5mm}
d: \hspace{0.5mm} Calar Alto $3.5$\,m telescope,\hspace{1.5mm}
e: \hspace{0.5mm} UV-Visual Echelle Spectrograph,\hspace{1.5mm}
f: \hspace{0.5mm} Space Telescope Imaging Spectrograph\hspace{1.5mm}\\
\vspace{-3mm}
}
\end{footnotesize}
\end{table*}

\begin{figure*}
  \resizebox{\hsize}{!}{\includegraphics{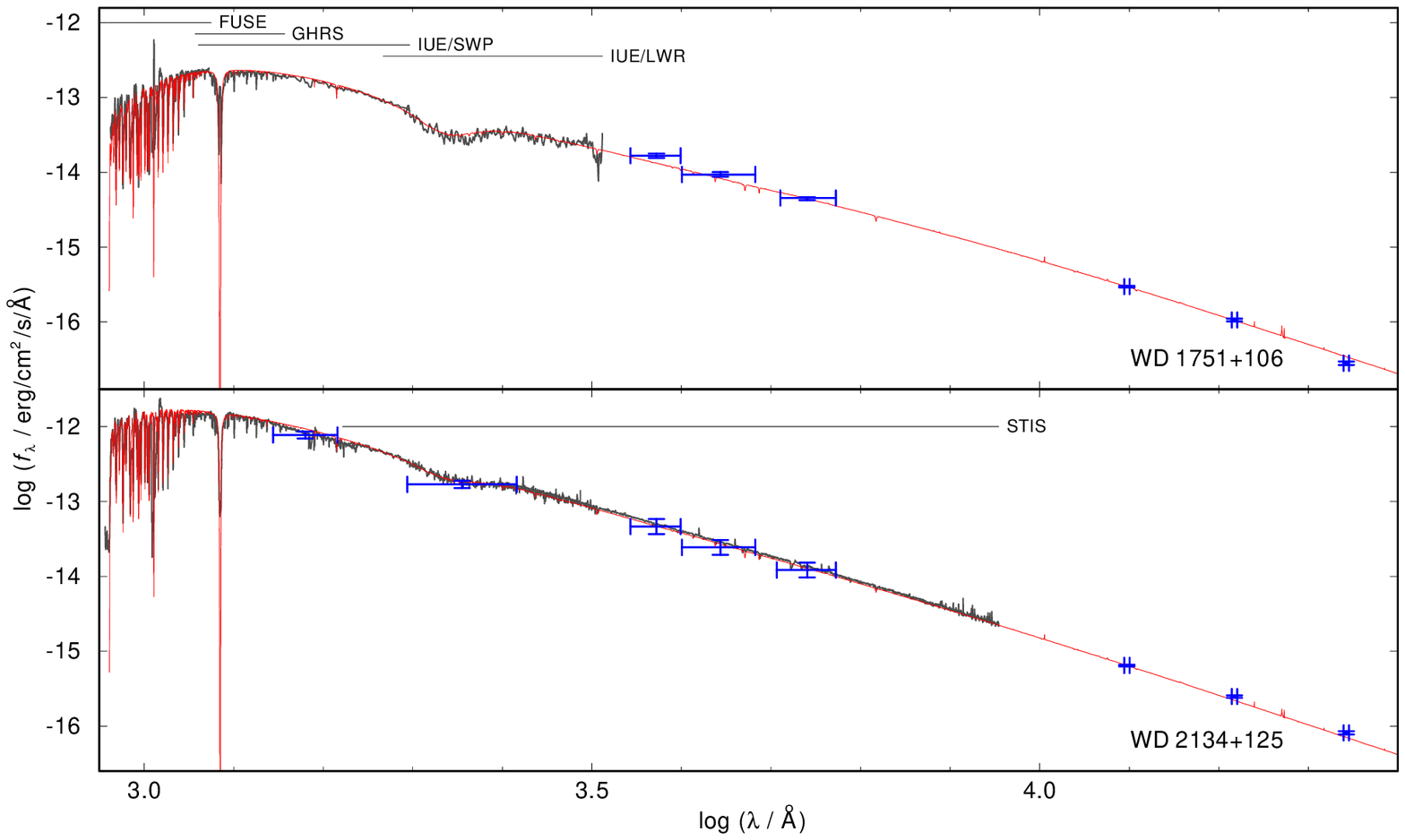}} 
   \caption[]{Determination of \ebv.
              Synthetic spectra of our best models (thick, red lines) of
              \wda (top panel, interstellar reddening with \ebvw{0.265} was applied) and
              \wdn (bottom, \ebvw{0.135}) 
               compared with the observations (black). The model fluxes are normalized to the 2MASS J magnitude \citep{cutrietal2003}.  
               U, B, and V magnitudes from \citet{ackeretal1992} and 
               GALEX FUV and NUV magnitudes (blue crosses) from \citet{binachietal2011} were added for \wdn  
              (flux conversion: \url{https://asd.gsfc.nasa.gov/archive/galex/FAQ/counts_background.html}). 
               Interstellar line absorption is included in the FUSE spectral range of the models.
             } 
   \label{fig:ebv}
\end{figure*}

\begin{figure*}
  \resizebox{\hsize}{!}{\includegraphics{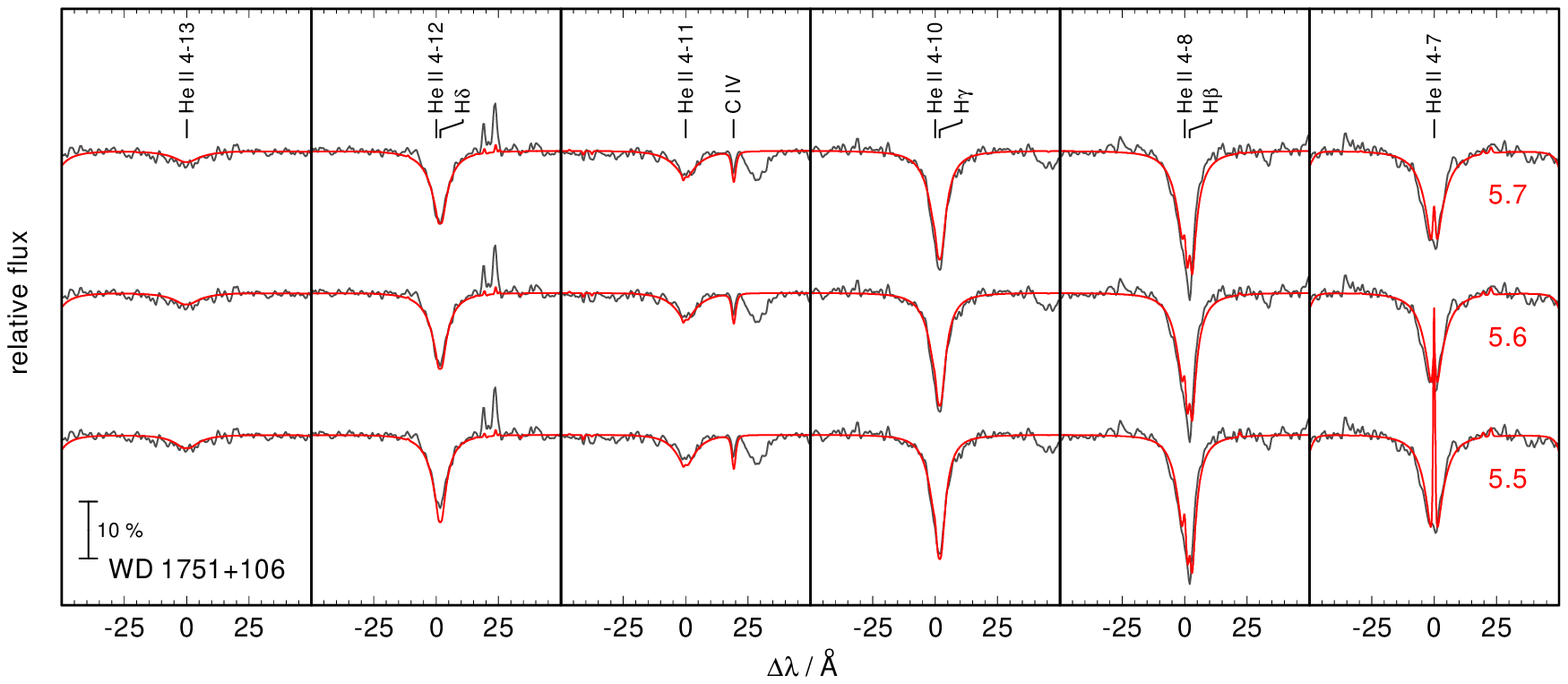}}
   \caption{Synthetic spectra calculated with \Teffw{115\,000} and different \logg, compared with the UVES SPY observations of 
            \ion{He}{ii} and \ion{H}{i} lines for \wda (gray).
            (\logg is indicated in panel 6)
            }
   \label{fig:logga43}
\end{figure*}
\begin{figure*}
  \resizebox{\hsize}{!}{\includegraphics{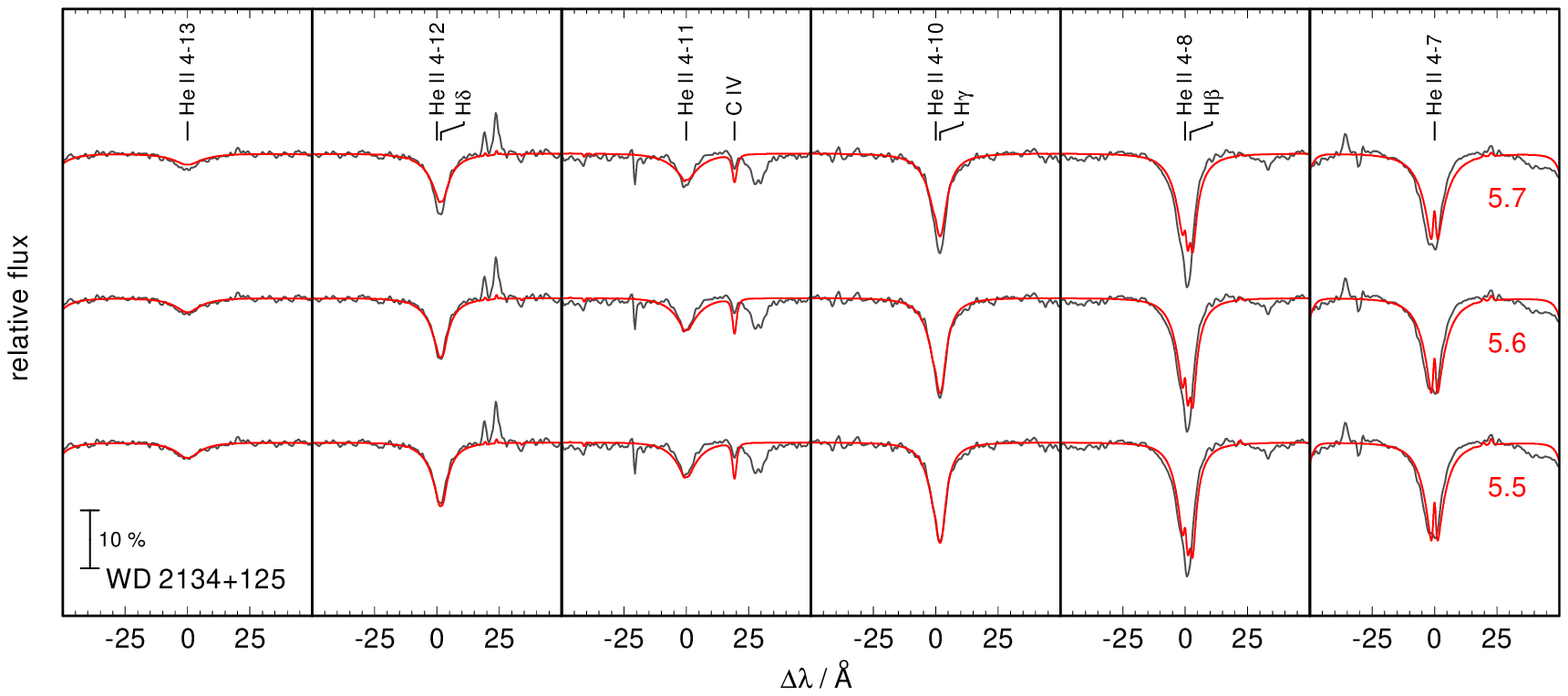}}
   \caption{Synthetic spectra calculated with \Teffw{115\,000} and different \logg, compared with the UVES SPY observations of 
            \ion{He}{ii} and \ion{H}{i} lines for \wdn (gray).
            (\logg is indicated in panel 6)
            }
   \label{fig:loggngc}
\end{figure*}
\begin{figure*}
  \resizebox{\hsize}{!}{\includegraphics{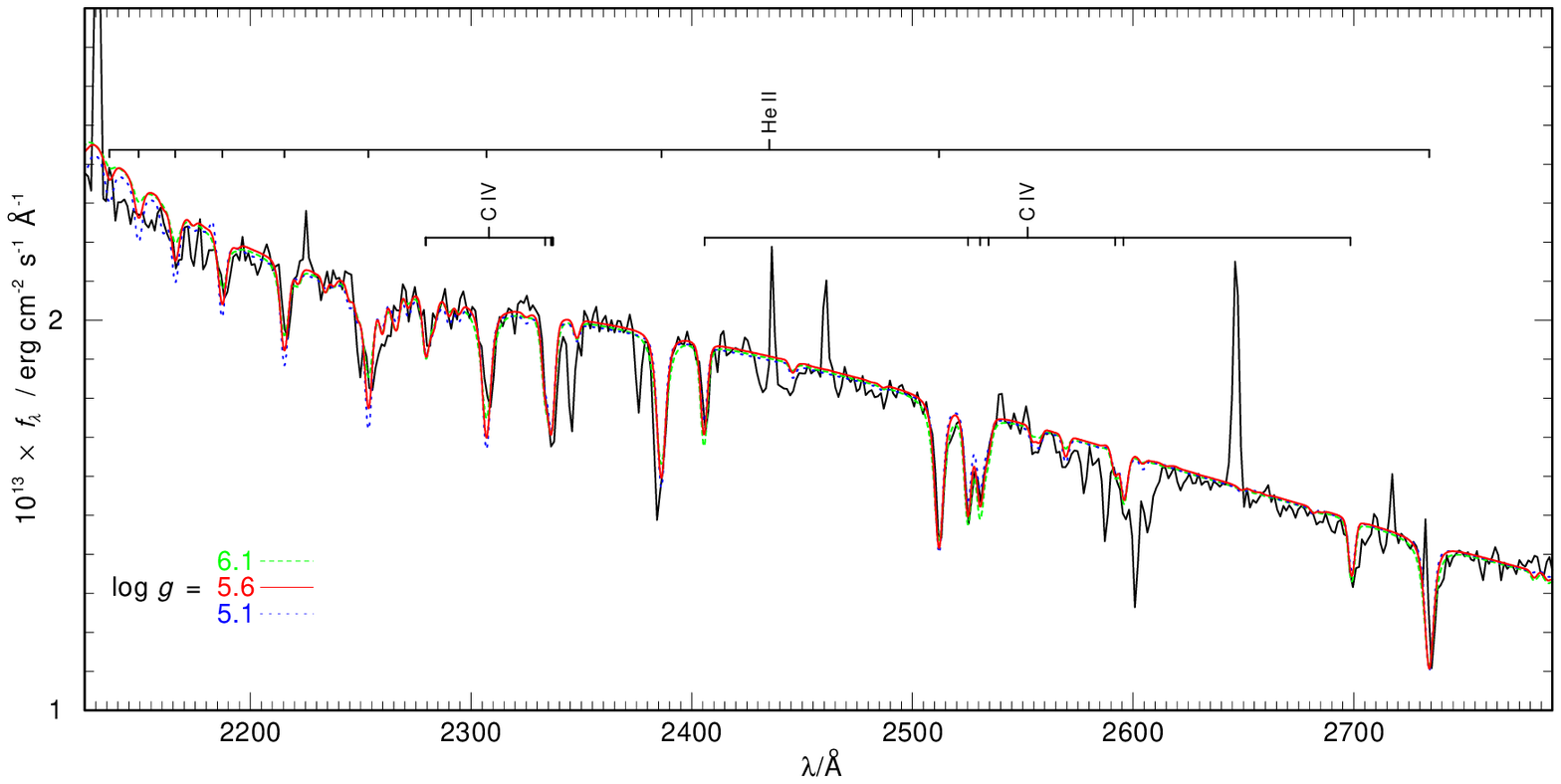}}
   \caption{Section of the HST/STIS spectrum of \wdn (black) compared with synthetic spectra calculated with \Teffw{115\,000} and different \logg of 6.1 (green dashed), 5.6 (red solid), and 5.1 (blue dotted).
            }
   \label{fig:fowlerngc}
\end{figure*}

\begin{figure*}
  \resizebox{\hsize}{!}{\includegraphics{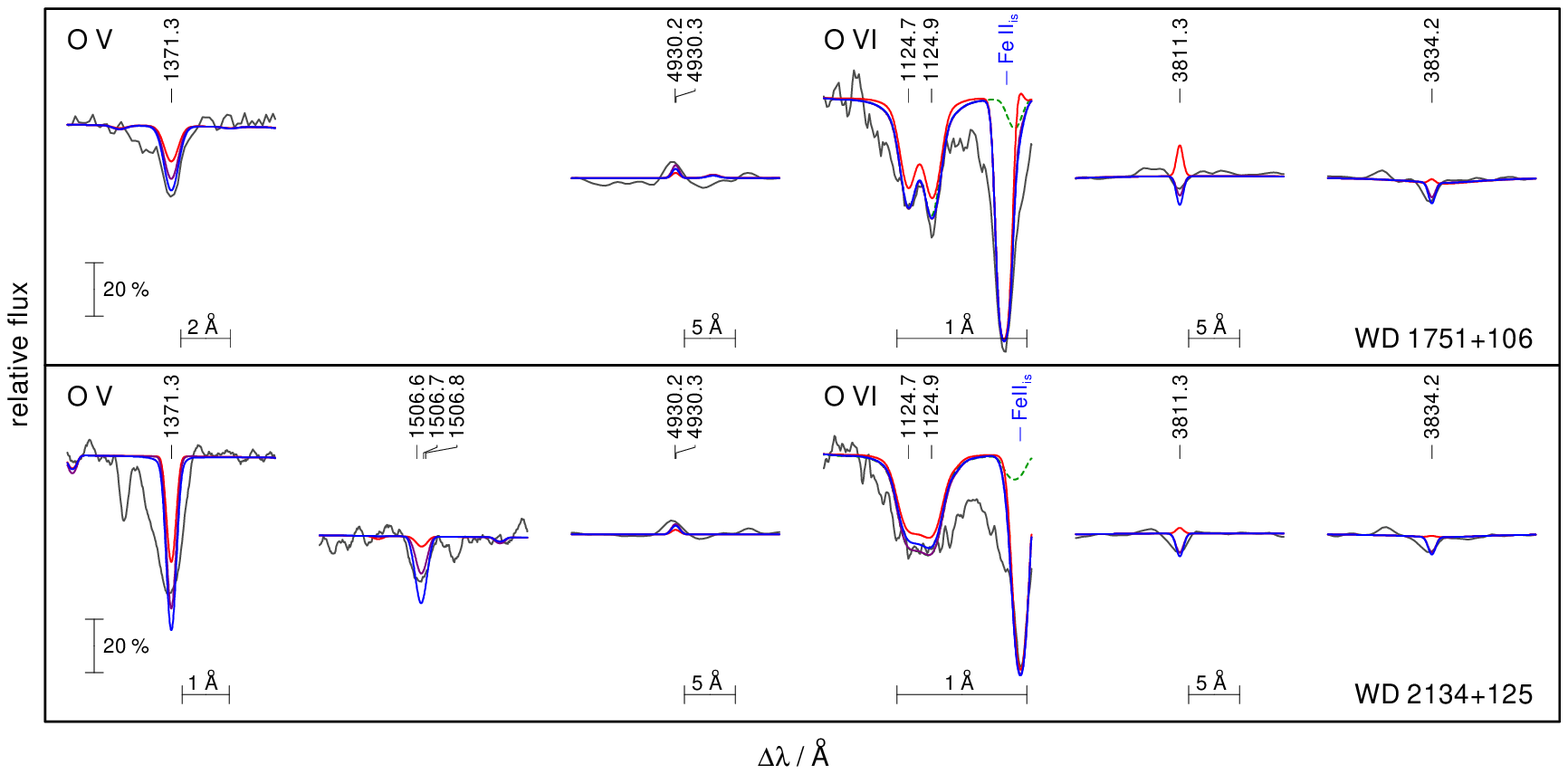}}
   \caption{Synthetic spectra calculated with \loggw{5.6} for \wda (top panel) and \wdn (bottom) 
     and different \Teff (red: \Teffw{125\,000}, purple: \Teffw{115\,000}, blue: \Teffw{105\,000}), compared with 
     the STIS observation of \Jonw{O}{v}{1371.3} and the FUSE observation of \Jonww{O}{vi}{1124.7, 1124.9} (gray). 
     A model without interstellar absorption (green, dashed) is shown in addition.
           }
   \label{fig:teffall}
\end{figure*}
\begin{figure*}
 \centering
   \caption[]{FUSE observation (gray) compared with the best static model including ISM line absorption (red). Stellar lines (black marks) and interstellar absorption features (blue) are identified at top. The green marks at the bottom of each panel indicate wavelengths of strong interstellar H$_2$ lines. This figure is available as online material of the published version only.
             } 
   \label{fig:fuseall}
\end{figure*}

\begin{figure*}
 \centering
   \caption[]{GHRS and STIS observation (gray) for \wda (top) and \wdn(bottom), respectively, compared with the best model (red). Stellar lines (black marks) and interstellar absorption features (blue) are identified at top. For wavelengths $>1450\,${\AA}, only the STIS spectrum is shown. This figure is available as online material of the published version only.
             } 
   \label{fig:stisall}
\end{figure*}

\begin{figure*}
  \resizebox{0.94\hsize}{!}{\includegraphics{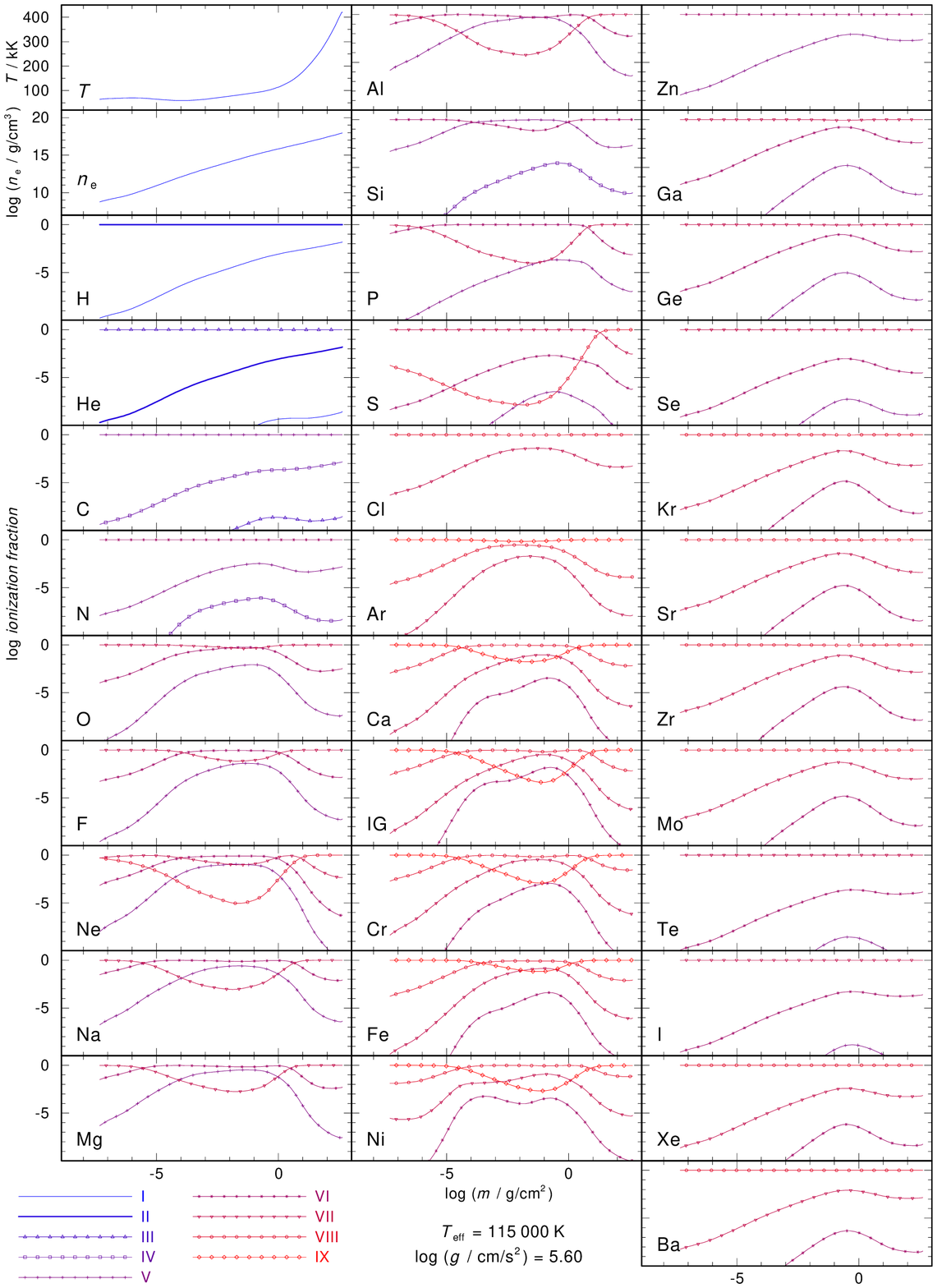}} 
   \caption[]{Temperature and density structures and ionization fractions of all ions which are considered in our final model for \wdn.
             } 
   \label{fig:ionfrac}
\end{figure*}

\begin{figure}
  \resizebox{\hsize}{!}{\includegraphics{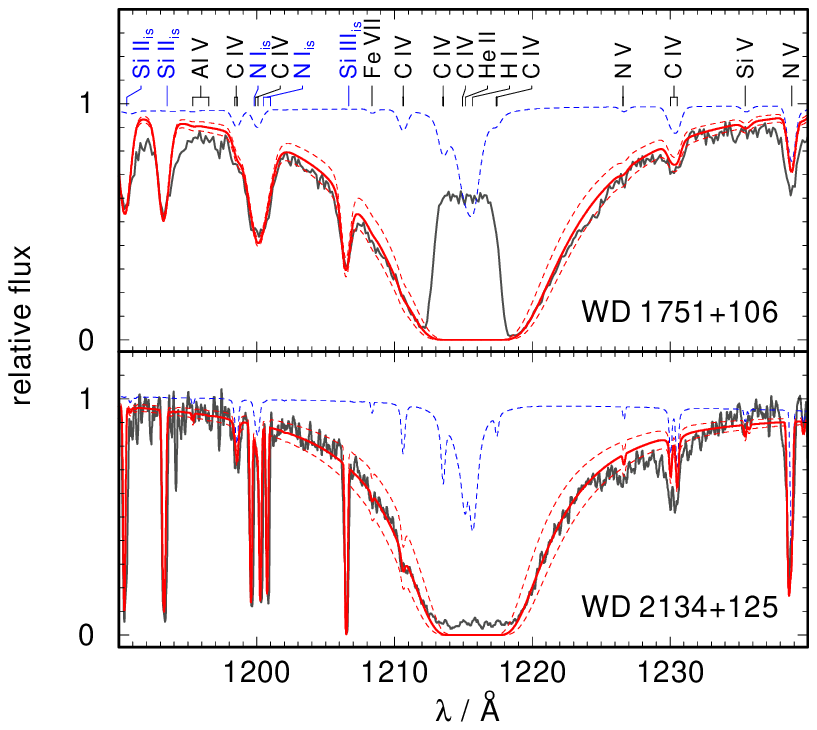}} 
   \caption[]{Synthetic spectra of our best models (red lines) for
              \wda (top panel) and
              \wdn (bottom) 
              around \Ion{H}{1} Ly\,$\alpha$ calculated with
              $n_\mathrm{H\,I} = 1.0 \times 10^{21}\,\mathrm{cm^{-2}}$ and
              $6.5 \times 10^{20}\,\mathrm{cm^{-2}}$, respectively,
              compared with the observations (gray). Models with $\Delta n_\mathrm{H\,I} = 2.0 \times 10^{20}\,\mathrm{cm^{-2}}$ 
              are shown (red, dashed).
              Spectra without interstellar absorption are shown for comparison (blue, dashed).
             } 
   \label{fig:nh}
\end{figure}
\begin{table}
\centering
\setlength{\tabcolsep}{0.3em}
\caption{Ions with recently calculated oscillator strengths.}
\label{tab:tee}
\begin{tabular}{rrll}
\hline
\hline
Zn & {\sc iv} &-\hspace{2mm}{\sc v}            & \citet{rauchetal2014zn} \\
Ga & {\sc iv} &-\hspace{2mm}{\sc vi}           & \citet{rauchetal2015ga} \\
Ge & {\sc  v} &-\hspace{2mm}{\sc vi}           & \citet{rauchetal2012ge} \\
Se & {\sc  v} &                                & \citet{rauchetal2017sesrtei} \\
Kr & {\sc iv} &-\hspace{2mm}{\sc vii}          & \citet{rauchetal2016kr} \\
Sr & {\sc iv} &-\hspace{2mm}{\sc vii}          & \citet{rauchetal2017sesrtei} \\
Zr & {\sc iv} &-\hspace{2mm}{\sc vii}          & \citet{rauchetal2016zr} \\
Mo & {\sc iv} &-\hspace{2mm}{\sc vii}          & \citet{rauchetal2016mo} \\
Te & {\sc vi} &                                & \citet{rauchetal2017sesrtei} \\
I  & {\sc vi} &                                &\citet{rauchetal2017sesrtei} \\
Xe & {\sc iv} &-\hspace{2mm}{\sc v}, {\sc vii} & \citet{rauchetal2015xe,rauchetal2016zr} \\
Ba & {\sc  v} &-\hspace{2mm}{\sc vii}          & \citet{rauchetal2014ba} \\
\hline
\end{tabular}
\end{table}

\begin{table}
\centering
\color{black}
\caption{Abundances used for the calculation of the atmospheric structures.}
\label{tab:pro2}
\begin{tabular}{rll}
\hline
\hline
 & \multicolumn{2}{c}{Mass fraction}\vspace{-0.17cm}\\
Element & &\vspace{-0.17cm}\\
 & \wda & \wdn\\
\hline
 H         & $2.5 \times 10^{-1}$ & $1.5 \times 10^{-1}$ \\
 He        & $4.6 \times 10^{-1}$ & $5.2 \times 10^{-1}$ \\
 C         & $2.7 \times 10^{-1}$ & $3.1 \times 10^{-1}$ \\
 N         & $2.6 \times 10^{-3}$ & $3.3 \times 10^{-4}$ \\
 O         & $4.4 \times 10^{-3}$ & $3.3 \times 10^{-3}$ \\
 F         & $3.3 \times 10^{-6}$ & $9.9 \times 10^{-6}$ \\
 Ne        & $1.2 \times 10^{-2}$ & $1.9 \times 10^{-2}$ \\
 Na        & $2.5 \times 10^{-5}$ & $2.4 \times 10^{-5}$ \\
 Mg        & $7.0 \times 10^{-4}$ & $5.9 \times 10^{-4}$ \\
 Al        & $1.2 \times 10^{-4}$ & $1.7 \times 10^{-4}$ \\
 Si        & $1.6 \times 10^{-4}$ & $1.2 \times 10^{-4}$ \\
 P         & $4.0 \times 10^{-6}$ & $4.9 \times 10^{-6}$ \\
 S         & $2.4 \times 10^{-4}$ & $5.0 \times 10^{-5}$ \\
 Cl        & $8.3 \times 10^{-6}$ & $8.1 \times 10^{-6}$ \\
 Ar        & $1.4 \times 10^{-4}$ & $3.5 \times 10^{-5}$ \\
 Ca        & $2.3 \times 10^{-5}$ & $1.0 \times 10^{-5}$ \\
 IG$^{(a)}$ & $1.8 \times 10^{-8}$ &$7.8 \times 10^{-9}$ \\
 Cr        & $6.0 \times 10^{-6}$ & $2.6 \times 10^{-6}$ \\
 Fe        & $4.5 \times 10^{-4}$ & $2.0 \times 10^{-4}$ \\
 Zn        & $6.5 \times 10^{-7}$ & $2.9 \times 10^{-7}$ \\
 Ga        & $2.0 \times 10^{-8}$ & $8.8 \times 10^{-9}$ \\
 Ge        & $8.5 \times 10^{-8}$ & $3.7 \times 10^{-8}$ \\
 Se        & $4.8 \times 10^{-8}$ & $2.1 \times 10^{-8}$ \\
 Kr        & $4.1 \times 10^{-8}$ & $1.8 \times 10^{-8}$ \\
 Sr        & $1.6 \times 10^{-8}$ & $7.1 \times 10^{-9}$ \\
 Zr        & $9.8 \times 10^{-9}$ & $4.3 \times 10^{-9}$ \\
 Mo        & $2.0 \times 10^{-9}$ & $8.7 \times 10^{-10}$\\
 Te        & $5.3 \times 10^{-9}$ & $2.3 \times 10^{-9}$ \\
 I         & $1.2 \times 10^{-9}$ & $5.4 \times 10^{-10}$\\
 Xe        & $6.3 \times 10^{-9}$ & $2.7 \times 10^{-9}$ \\
 Ba        & $6.7 \times 10^{-9}$ & $2.9 \times 10^{-9}$ \\
\hline
\end{tabular}\\
\color{black}
\textbf{Notes.} $^{(a)}${IG is a generic model atom \citep[cf., ][]{rauchdeetjen2003} that includes opacities of Sc, Ti, V, Mn, Ni, and Co.}
\end{table}


\bsp	
\label{lastpage}
\end{document}